\documentclass{mn2e}
\usepackage[pass]{geometry}
\usepackage{newtxtext,newtxmath}
\usepackage{pdflscape}
\usepackage{graphicx}
\usepackage{amsmath}
\usepackage{times}

\def\arcdeg{\hbox{$^\circ$}}
\def\arcsec{\hbox{$^{\prime\prime}$}}
\def\deg2{\hbox{$\rm deg^{2}$}}

\def\lsim{\mathrel{\rlap{\lower4pt\hbox{\hskip1pt$\sim$}}\raise1pt\hbox{$<$}}} 
\def\gsim{\mathrel{\rlap{\lower4pt\hbox{\hskip1pt$\sim$}}\raise1pt\hbox{$>$}}} 

\begin{document}

\title[Spectroscopic Classifications]{Spectroscopic Classification of Extragalactic Transients from CRTS}
\author[A.J. Drake et al.]{
A.J.~Drake,$^1$ S.G.~Djorgovski,$^1$ M.J.~Graham,$^1$ D.~Stern,$^2$ A.A.~Mahabal,$^1$ 
\newauthor M.~Catelan,$^{3,4}$ J.L.~Prieto,$^{4,5}$ and S.~Larson$^6$ 
\\
$^1$California Institute of Technology, 1200 E. California Blvd, CA 91225, USA\\
$^2$Jet Propulsion Laboratory, California Institute of Technology, 4800 Oak Grove Drive, MS 169-221, Pasadena, CA 91109, USA\\
$^3$Pontificia Universidad Cat\'olica de Chile, Instituto de Astrof\'isica, 
Facultad de F\'{i}sica, Av. Vicu\~na Mackena 4860, 7820436 Macul, Santiago, Chile\\
$^4$Millennium Institute of Astrophysics, Nuncio Monse{\~n}or S{\'o}tero Sanz 100, Off. 104, Providencia, Santiago, Chile\\
$^5$Instituto de Estudios Astrof\'isicos, Facultad de Ingenier\'ia Ciencias, Universidad Diego Portales, 
Av. Ej\'ercito Libertador 441, Santiago, Chile\\
$^6$The University of Arizona, Department of Planetary Sciences,  Lunar and Planetary Laboratory, 
1629 E. University Blvd, Tucson AZ 85721, USA\\
}

\volume{000}
\pubyear{0000}
\maketitle

\begin{abstract}
\noindent 
The Catalina Real-time Transient Survey (CRTS) carried out a public survey for optical
transients between 2007 and 2019, discovering more than 16,000 transient candidates.
Here we present the spectra and highlight the results of the spectroscopic follow-up
of CRTS extragalactic transients. As expected, we find that the bulk of these transients
are normal supernovae. However, as we prioritised transients exhibiting unusual features
or environments during our spectroscopic follow-up, we focus on the rarer types of transients.
These objects include more than a dozen type-I superluminous supernovae and dozens of
type-I and II supernovae that underwent circumstellar medium interactions.
We highlight several specific supernovae, including a new analysis of SN 2008iy, a type-IIn
which exhibited a bright pre-supernova outburst event similar to SN 2009ip and lasted more
than 1800 days; CSS111225:140122+161705, a type-I supernova that showed an extreme 2.5 magnitude
rebrightening event more than 200 days after its initial outburst; and SN 2009ny, a type-Ibn
supernova that exhibited strong helium emission lines similar to SN 2002ao. We confirm our
previous finding that numerous CRTS transients are associated with galaxies of extremely
low luminosity. We discuss the difficulty in determining the origin of transients associated
with outbursts in active galactic nuclei (AGN), tidal disruption events, and type-IIn supernovae.
As an example, we present CSS150120:110008+385352, a CRTS transient similar to CSS100217:102913+404220
that occurred within a quiescent AGN and peaked at $M_V = -23.6$.
\end{abstract}

\section{Introduction}

Our understanding of the nature of astronomical transient variability has evolved significantly over 
the past two decades.  Initial progress in this endeavour has been largely due to wide-field transient 
surveys, such as the Catalina Real-time Transient Survey (CRTS, Drake et al.~2009a\nocite{Dra09a}), 
Djorgovski et al.~2012\nocite{Djo12}),
the Panoramic Survey Telescope and Rapid Response System (PanSTARRS, Chambers et al.~2016\nocite{Cha16}), and the 
Palomar Transient Factory (PTF, Law et al.~2009\nocite{Law09}). More recently, the Asteroid Terrestrial-impact Last 
Alert System (ATLAS, Tonry et al.~2018\nocite{Ton18}) and the Zwicky Transient Facility (ZTF, Bellm et al.~2019\nocite{Bel19}) 
surveys have continued to expand our understanding with much faster
coverage, while much deeper transient surveys, such as the Vera C Rubin's Large Survey of Space and 
Time \cite{Ive19} and the surveys to be conducted with the Nancy Grace Roman (Akeson et al.~2019\nocite{Ake19}; 
Schileder et al.~2024\nocite{Sch24}), are set to come online within the next few years. 

While many types and timescales of variability have been detected in transient surveys, the most 
common source of bright ($M_{V} < -14$) events that are truly transient, are supernovae (SNe).
Supernovae provide insight into the final stages of stellar evolution, and, in the case of type-Ia 
supernovae, supply uniform standard candles for measuring cosmological distances \cite{San82}.

Over the last decade, supernova discovery rates have risen from hundreds to thousands of events per year. 
This increase is almost exclusively due to the advent of modern transient surveys and has been so dramatic 
that it outstrips our ability to spectroscopically confirm the nature of most transients.
For instance, between 2007 and 2019, CRTS found more than 16,000 transients and highly variable sources. 
Of these, more than 4000 of the transients were classified as supernovae. Yet without dedicated follow-up,
only a fraction of these supernova candidates could be spectroscopically confirmed.
Even surveys with dedicated follow-up using specialized robot low-resolution instruments, such the Spectral 
Energy Distribution Machine (SEDM, Blagorodnova et al.~2018\nocite{Bla18}), operated by ZTF \cite{Bel19},
remains insufficient to characterize all of the ever-growing numbers of supernova discoveries.

Although supernovae are the most common source of bright extragalactic optical transients, many galaxies 
harbour massive black holes that undergo accretion processes that are often observed as nuclear variability.
The variations of these sources are usually a few tenths of a magnitude but can become compounded to 
variations of more than a magnitude on long timescales (MacLeod et al.~2012\nocite{Mac12}, Graham et al.~2017\nocite{Gra17}). 
In very rare cases, these Active Galactic Nuclei (AGN) have recently been found that appear to turn on 
or off over timescales of years (MacLeod et al.~2016\nocite{Mac16}, Gezari et al.~2017\nocite{Gez17}, Stern et al.~2018\nocite{Ste18}).
This so-called ``changing-look'' behaviour has been observed at low levels in target studies of Seyfert
galaxies for decades (e.g. Tohline \& Osterbrock 1976\nocite{Toh76}; Osterbrock \& Shuder 1982\nocite{Ost82}; Goodrich 1989\nocite{Goo89}) but 
in the last decade has only been observed in luminous quasars. The advent of large surveys has confirmed this as 
a widespread phenomenon. Additionally, with the increased monitoring and improved selection techniques,
large numbers of Tidal Disruption Events (TDEs) are now routinely discovered by transient 
surveys (e.g. van Velzen et al. 2021\nocite{van21}, Hammerstein et al.~2023\nocite{Ham23}).

Rarer types of luminous extragalactic optical transients are also being detected by transient surveys in 
increasing numbers as their surveying power has increased. These include Novae, Luminous Red Novae 
(LRN, Pastorello et al.~2019\nocite{Pas19}), gamma ray burst (GRB) counterparts \cite{Sah97},
Fast Blue Optical Transients (FBOTS, Coppejans et al.~2020\nocite{Cop20}), and Kilonovae (Smartt et al.~2017\nocite{Sma17}).

Within CRTS we specifically targeted our spectroscopic observations on transients that appeared 
to have different properties than the bulk population. The targets included: transients without any 
obvious host galaxy or stellar association; events where a host redshift or a long timescale suggested 
and event was superluminous; and transients that were detected within hours of our follow-up.
In this paper, we investigate the nature of the resulting discoveries. 
We adopt $H_{0}\rm = 72\,\,km\,\,s^{-1} Mpc ^{-1}$, $\Omega_\Lambda = 0.73$, 
$\rm \Omega_M = 0.27$ and use Vega magnitudes throughout.

\section{The CRTS Survey}

The main goal of CRTS\footnote{http://crts.caltech.edu/} was the rapid discovery and public announcement 
of transients and highly variable ($>1$ magnitude) sources. CRTS began operation in late 2007 
(Drake et al.~2009a\nocite{Dra09a})
using observations taken by the Catalina Sky Survey\footnote{http://www.lpl.arizona.edu/css/} 
Near-Earth Object search project \cite{Lar03}.
The Catalina Sky Survey initially consisted of three 
telescopes: the 0.7m Catalina Schmidt Survey (CSS\footnote{Note: In this work CSS relates to the telescope rather 
than the survey.}) and the 1.5m Mount Lemmon Survey (MLS) telescopes, both in Tucson, Arizona; 
and the 0.5m Siding Spring Survey (SSS) telescope at Siding Spring, Australia. 

All of the observations made with the Catalina telescopes and analyzed by CRTS were taken using 
fixed fields. This allowed new data in a specific location to be compared directly with prior observations. 
The combined dataset from the three telescopes covered most of the sky between declinations $\delta = -75$ and +65 degrees. 
However, regions within $\sim$15$\arcdeg$ of the Galactic plane were generally avoided due to crowding. 
The MLS, SSS and CSS $\rm 4k^2$ pixel CCD cameras initially covered 1.2 $\rm deg^2$, 4.0 $\rm deg^2$ 
and 8.2 $\rm deg^2$, respectively. The MLS and CSS were upgraded to $\rm 10.5k^2$ cameras, in 2016 and 2017, 
increasing their coverage to 5 $\rm deg^2$ and $\sim$20 $\rm deg^2$, respectively. However, CRTS only 
searched for transients in the CSS images taken between September 2007 and October 2017; the MLS images 
taken between October 2009 and February 2019; and SSS observations between May 2010 and July 2013.

In most cases, the observations were taken in sequences of four images separated by 10 minutes.  
All of the images were unfiltered. Isophotal photometry was obtained for all sources in each
observation using the SExtractor photometry program parameter ``MAG\_ISOCOR'' \cite{Ber96}.
The extracted photometry was calibrated to a pseudo-$V$ magnitude ($\rm V_{CSS}$) using a few dozen pre-selected 
standard stars in each field. Further details of the photometric calibration and transformations to standard systems 
are given in Drake et al.~(2013a)\nocite{Dra13a}. The photometry taken with the original survey cameras 
is publicly available as part of Catalina Surveys Data Releases\footnote{http://nesssi.cacr.caltech.edu/DataRelease},
while more recent photometry will be available in a future data release.

\section{Optical Transients}

\subsection{Detection}

Transient searches within CRTS were carried out by identifying new sources detected in multiple 
images of an observing sequence, or previously detected sources exhibiting significant brightness 
variations. 
However, unlike newer synoptic surveys, CRTS mainly concentrated on detecting high-amplitude transient 
events. Thus, CRTS specifically filtered the detections of hundreds of thousands of low-amplitude 
variable sources (due to common variable stars and AGN). This was achieved by using a threshold 
of $\gsim$ 1 magnitude if a detection matched an existing source, or $>3 \sigma$ threshold 
when no prior detections were present. This threshold was reduced from an initial value 2 magnitudes 
(Drake et al.~2009a\nocite{Dra09a}) as it was found to be sufficient for removing non-transient 
sources, including most periodic variables, while still including outbursting objects, such as 
cataclysmic variables (CVs), AGN and blazars. In all cases, we required that detections were made 
in at least three of the four images in a sequence. This detection coincidence requirement removed 
most spurious detections due to image artifacts.

To remove spurious transient candidates due to real astronomical objects, such as asteroids, 
we automatically compared each source detection to known asteroid locations calculated from 
Minor Planet Center (MPC) catalogues. However, since asteroid catalogues were incomplete at
the depth of the images when the survey began, we used the astrometry from the individual 
detections to remove any moving objects. 

In general, we were able to automatically extract transient candidates within 10 minutes 
of observation. The CRTS processing pipeline also automatically extracted all the archival 
photometry and image cutouts at the location of every candidate. All of the transient candidates 
passing the initial filtering process were immediately published via Public Virtual Observatory Event 
(VOEvent, Seaman et al.~2011\nocite{Sea11}) alert notifications on VOEventNet \cite{Dra06}. 
Additionally, once a candidate was human-verified, any future observations were automatically 
added to the lightcurve along with the associated meta-data.

\subsection{Source Classification}

The classification of transients and variable sources is a complex process. 
Within CRTS the initial step primarily consisted of comparing sets of flux measurements 
along with spatial information. In most cases, a rough classification simply required 
combining new information with historical CRTS lightcurves and archival multi-wavelength 
catalogues or images. The classification could often be significantly improved by matching 
transient detection locations with hundreds of past survey images, catalogues, and 
references using the National Virtual Observatory (NVO) Datascope service \cite{McG08}. 

For example, most CRTS supernovae within the footprint of the Sloan Digital Sky Survey
(SDSS, Abazajian et al.~2009\nocite{Aba09}) were clearly associated with galaxies seen in the
SDSS images (which are deeper and have higher resolution than CRTS images). Likewise, the UV Ceti
flares generally match faint red stars in SDSS, and most CRTS CVs match faint blue stars
(see Drake et al.~2014a\nocite{Dra14a}). 
However, since CRTS covers more area than SDSS, data from 2MASS (Two Micron All Sky Survey, 
Skrutskie et al.~2006\nocite{Skr06}) and GALEX (Galaxy Evolution Explorer, Martin et al.~2005\nocite{Mar05}) 
were also used to classify CVs and flare stars. In contrast, many objects exhibiting repeated
bright flaring activity within their lightcurves match known radio sources in surveys, such as
FIRST (Faint Images of the Radio Sky at Twenty-Centimeters, White et al.~1997\nocite{Whi97}) 
and the NVSS (NRAO VLS Sky Survey, Condon et al.~1998\nocite{Con98}), and were clearly associated
with blazars. In the case of regular radio-quiet AGN, the sources generally had much less variability.
In some cases, classifications was clear due to prior identifications, or other existing public
information (such as SDSS spectra).

To aid our classification of CRTS transients we obtained $g$, $r$, $i$ and $z$-band follow-up photometry 
with the Palomar $60^{\prime\prime}$ (P60) telescope between December 2007 and December 2013. In total, 
we took multi-band photometry for 1765 of the CRTS transients to a depth of $> 21$. For about 10\%
of the transients we obtianed multiple nights of photometry to improve our classifications using colour
or brightness evolution.

In total the CRTS pipeline automatically triggered on 160,000 CRTS transient candidates and variables
during 4860 nights of observations from all three CSS telescopes. Of these, $\sim10$\% were genuine
transients or variable sources, and $\sim10$\% were redetections of the same candidates. All of the
automated detections were reviewed and classified by a single person (AJD) to maintain consistency
within our selection process. In most cases, these human-classifications were performed within 12 hours
of detection. Each classification was assigned a rough level of certainty based on the results
of our photometric and spectroscopic follow-up campaigns. More than a thousand of the CRTS SN
candidates were posted in ATels\footnote{https://www.astronomerstelegram.org/} to maximize the
scientific utility of the events.

Beiginning in October 2010 we started posting CRTS classification circulars to provide more detailed
information for all vetted transient candidates. These circulars were distributed as VOEvents that
were linked to each candidate transient. Each notice included the context information used in the
classification process. For instance, positional matches to known SDSS galaxies, or detections in
optical, UV, near-IR, or radio surveys, as well as information about prior detections in CRTS data.
Additionally, in 2011 we began submitting bright CRTS supernova candidates to the CBAT 
(Central Bureau for Astronomical Telegrams) Transient Objects Confirmation Page 
(TOCP)\footnote{http://www.cbat.eps.harvard.edu/unconf/tocp.html}.

While the CRTS event classifications were found to be generally accurate, the process was
subject to human selection biases. Transient surveys such as ZTF have continued to automate 
the classification process by using multiple machine-learning techniques \cite{Mah19}.
This enables them to provide results in a uniform manner that can be easily modified if
errors are found with the results. The use of an automated process also allows them to classify
far larger numbers of candidates. As a result, new transient surveys can reduce their detection
thresholds and process millions of variable sources daily \cite{Pat19}.
Recent work has also proceeded to pass the classification and distribution tasks to alert event brokers, 
such as ALeRCE (Automatic Learning for Rapid Classification of Events, Forster et al.~2021\nocite{For21}) 
and Lasair \cite{Smi19a}, 
that ingesting event metadata and provide rapid, uniform classifications for transient candidates.
Nevertheless, machine-learning based classification methods still face difficulties when they
encounter new types of objects.

\subsection{Follow-up Target Selection}

All CRTS transient candidate detections were immediately public, and classifications were 
almost always public within a day. Thus, obtaining spectroscopic follow-up for every object was not 
the top priority. Rather, to maximize potential utility to the astronomical community, we generally 
prioritized our spectroscopic confirmation to transients that were either newly discovered and likely
rising, or bright for an extended period. Additionally, since the follow-up of nearby type-Ia 
supernovae was already the main goal of many surveys (e.g. The Supernova Factory (SNF), Copin et al.~2009\nocite{Cop09};
Lick Observatory Supernova Survey (LOSS), Graur et al.~2017\nocite{Grau17}; The Carnegie Supernova Project (CSP), 
Hsiao et al.~2019; Center force Astrophysics-4 (CfA4), Hicken et al.~2012\nocite{Hic12}), we attempted to limit 
any duplication by prioritizing other types of events. 

The resulting optical follow-up spectra largely consist of a small number of high amplitude outbursts
of variable stars, such as CVs \cite{Dra14a}, flare stars (Mahabal et al. in prep.), blazars 
(Djorgovski et al. in prep.), and supernovae (this work). Nevertheless, even after deprioritizing
likely SN-Ia, the rate of transient detection still far outstripped available spectroscopic observing
time. Thus, we prioritized confirmations of objects that were brighter than 20th magnitude, with
events occurring in the faint host galaxies given higher priority. We aimed to better understand 
the extremes of the transient population, rather than concentrate on the common types.

\subsection{Spectroscopic Data}



During our spectroscopic follow-up campaigns, we obtained 353 spectra for 301 unique extragalactic
transient candidates. These observations were taken on 112 nights spanning October 2009 to January 2019.
Among them, 268 were observed with the Palomar-5m telescope using the Double Beam Spectrograph (DBSP,

\newpage
\begin{landscape}
\begin{table}
\caption{CRTS transient properties and spectroscopic identifications}
\begin{tabular}{lllrcllccr}
\hline
CRTS ID & RA (J2000) & Dec & $z$ & $V_{\rm max}$ & Class & Instrument & Date Obs & Alt ID & Refs\\ 
 \hline
CSS080312:102245+021753 & 10:22:44.65 & +02:17:52.5 & 0.100 & 18.82 &  IIn & Keck+LRIS & 2009-12-20 & - & - \\
CSS080701:234413+075224 & 23:44:12.81 & +07:52:23.5 & 0.068 & 18.37 &  IIn & Keck+LRIS & 2010-10-04 &  2008ja &  C1936\\
 &  &  &  &  &  & P200+DBSP & 2009-07-22 & & \\
CSS080921:012706+031306 & 01:27:05.78 & +03:13:05.7 & 0.063 & 17.74 &  Ia & Keck+LRIS & 2009-12-20 &  2008gs &  C1559\\
CSS080922:231617+114248 & 23:16:16.59 & +11:42:47.5 & 0.133 & 16.89 &  SLSN-IIn & P200+DBSP & 2009-06-22 &  2008fz & A1734, A1778, C1524, C1533\\
CSS080928:160837+041627 & 16:08:37.22 & +04:16:26.6 & 0.0407 & 17.76 &  IIn & Keck+LRIS & 2011-09-01 &  2008iy &  C1780\\
 &  & & & & & Keck+LRIS & 2013-06-10 & & \\
 &  & & & & & P200+DBSP & 2009-07-22 & & \\
 &  & & & & & P200+DBSP & 2010-03-15 & & \\
 &  & & & & & P200+DBSP & 2010-06-11 & & \\
 &  & & & & & P200+DBSP & 2010-09-14 & & \\
 &  & & & & & SMARTS+RCspec & 2009-03-27 & & \\
 &  & & & & & SMARTS+RCspec & 2009-04-20 & & \\
CSS081001:003705-060939 & 00:37:04.77 & $-$06:09:39.2 & 0.089 & 18.37 &  Ia & P200+DBSP & 2008-10-01 &  2008ix &  C1766\\
CSS081009:002151-163204 & 00:21:51.39 & $-$16:32:03.7 & 0.285 & 18.52 &  SLSN-I? & Keck+LRIS & 2009-12-20 & - & A1823, A1833 \\
CSS081030:043655-002136 & 04:36:55.19 & $-$00:21:35.6 & 0.130 & 18.10 &  IIn/Ia-CSM & Keck+LRIS & 2009-12-20 &  2008iu &  C1681\\
 &  & & & & & SMARTS+RCspec & 2008-12-04 & & \\
CSS081201:103354-032125 & 10:33:53.79 & $-$03:21:25.3 & 0.0709 & 18.18 &  IIn & Keck+LRIS & 2009-12-20 &  K0812-2 &  C1682\\
 &  & & & & & P200+DBSP & 2009-04-01 & & \\
CSS081220:024813+211047 & 02:48:13.27 & +21:10:47.2 & 0.220 & 19.19 &  SLSN-IIn & Keck+LRIS & 2009-12-20 & - & - \\
CSS090102:130037+175057 & 13:00:37.49 & +17:50:57.0 & 0.143 & 18.93 &  Ic & P200+DBSP & 2009-02-25 &  2009de &  C1766\\
CSS090131:144115-035453 & 14:41:14.85 & $-$03:54:53.4 & 0.0715 & 18.74 &  IIn & Gemini+GMOS & 2010-01-16 & - & - \\
CSS090213:030920+160505 & 03:09:19.78 & +16:05:05.3 & 0.0315 & 16.89 &  Ia & Keck+LRIS & 2009-12-20 &  2009aq & A1950, C1713\\
 &  & & & & & P200+DBSP & 2009-02-25 & & \\
CSS090216:100910+075434 & 10:09:10.10 & +07:54:34.4 & 0.069 & 18.61 &  Ia-CSM? & P200+DBSP & 2009-04-01 &  2009bx & A1937, C1744, C1766\\
CSS090219:095526-012821 & 09:55:25.70 & $-$01:28:21.2 & 0.025 & 18.36 &  IIb & P200+DBSP & 2009-02-25 &  2009ar &  C1713\\
CSS090301:160553+172446 & 16:05:52.66 & +17:24:45.5 & 0.118 & 18.58 &  Ia & P200+DBSP & 2009-04-01 &  2009df &  C1766\\
CSS090303:113345+143448 & 11:33:44.93 & +14:34:47.8 & 0.094 & 18.64 &  Ia & P200+DBSP & 2009-04-01 &  2009dg &  C1766\\
CSS090303:135819+201051 & 13:58:19.37 & +20:10:51.1 & 0.055 & 18.78 &  IIP & P200+DBSP & 2009-04-01 &  2009dh &  C1766\\
CSS090318:094055+011608 & 09:40:55.44 & +01:16:08.2 & 0.018 & 16.90 &  IIP & SMARTS+RCspec & 2009-03-26 &  2009cy  &  C1755\\
CSS090321:113658+450051 & 11:36:57.86 & +45:00:50.7 & 0.134 & 18.66 &  Ic & P200+DBSP & 2009-04-01 &  2009di &  C1766\\
CSS090324:135835+311252 & 13:58:34.74 & +31:12:52.3 & 0.116 & 19.12 &  Ia & P200+DBSP & 2009-04-01 &  2009dj &  C1766\\
CSS090401:113024+021435 & 11:30:23.63 & +02:14:35.1 & 0.069 & 19.41 &  Ia & P200+DBSP & 2009-04-01 &  2009dk &  C1766\\
CSS090417:133946-212719 & 13:39:46.36 & $-$21:27:19.0 & 0.062 & 19.36 &  Ia-91bg & P200+DBSP & 2009-04-28 &  2009du &  C1791\\
CSS090418:132440+163405 & 13:24:40.14 & +16:34:05.4 & 0.029 & 19.25 &  IIP & P200+DBSP & 2009-04-28 &  2009dv &  C1791\\
CSS090421:164926+055248 & 16:49:25.66 & +05:52:48.0 & 0.046 & 17.51 &  Ia & P200+DBSP & 2009-04-28 &  2009dx &  C1791\\
CSS090422:150104+431314 & 15:01:04.03 & +43:13:13.9 & 0.094 & 18.41 &  Ia & P200+DBSP & 2009-04-28 &  2009dy &  C1791\\
CSS090516:104522+105717 & 10:45:22.17 & +10:57:17.0 & 0.243 & 18.30 &  SLSN-I? & Keck+LRIS & 2010-01-07 &  2009fp & A2057, C1825\\
CSS090529:163854+214906 & 16:38:54.17 & +21:49:05.9 & 0.047 & 18.12 &  IIP & P200+DBSP & 2009-06-09 &  2009go &  C1861\\
\hline
\end{tabular}
\medskip\\
The full table is available in the online version.\\
Col. (1) presents the CRTS identifier.\\
Cols. (2) \& (3), present the right ascension and declination, respectively.\\
Col. (4) presents the redshift.\\
Col. (5) presents the observed peak $V$ magnitude.\\
Col. (6) presents the classification. Uncertain classifications are denoted with ``?''.\\
Col. (7) presents the telescope and spectrograph used in the observation.\\
Col. (8) presents the observation date.\\
Col. (9) presents alternate identifiers (such as IAU ID), if any.\\
Col. (10) presents the event notification identifier (if any). ``C'' denotes CBETs, and ``A'' denotes ATels. Numbers denote the article number.\\
\end{table}
\end{landscape}

\noindent
Oke \& Gunn~1982\nocite{Oke82}), 53 with the Keck-10m telescope using the Low Resolution Imaging 
Spectrometer (LRIS, Oke et al.~1995\nocite{Oke95}), 20 with the Gemini-8m North and South telescopes
with the Gemini Multi-Object Spectrographs (GMOS, Hook et al.~2004\nocite{Hoo04}), and 12 with the
Small and Medium Aperture Research Telescope System 
(SMARTS, Subasavage et al.~2010\nocite{Sub10}) using the 1.5m RC spectrograph. 
All of the spectra were reduced with standard IRAF processing software and methods. 
The spectroscopic data are sampled at scales of between 0.6 and 2.2 \r{A}/pixel 
and have a typical resolving power of $R=1000$. In most cases, the signal-to-noise ratio of the spectra
is more than sufficient in determining the presence of the general features of supernovae and other
optical transients. Of the inital spectra 321 were successfully reduced and found to be likely
extra-galactic candidates. Nevertheless, the classification remain ambiguous in a few cases. 

In Table 1, we present the details of the CRTS extragalactic transient spectroscopy, including
IDs\footnote{The CRTS identifiers provided consist of the source discovery date followed by 
the approximate right ascension and declination of the event.}, observational parameters, 
redshifts, classifications and references to discovery notices. The discovery notices generally 
consisted of Astronomers Telegrams (ATels) \& Central Bureau Electronic Telegrams (CBETs).

\section{Supernovae}

The most common transient events discovered by CRTS are supernovae. These include the general classes
of SN-Ia, SN-II, superluminous supernovae (SLSNe), and large numbers of supernovae that exhibit emission
features due to interaction shocks in their circumstellar media (e.g. SN-IIn). For most SN classifications,
we rely on the Supernova Identification (SNID) software of Blondin \& Tonry (2007)\nocite{Blo07}. However, 
in a some cases automated classification was not possible and it was necessary to compare the spectral features.

\subsection{Type-Ia Supernovae}

\begin{figure}{
\includegraphics[width=84mm]{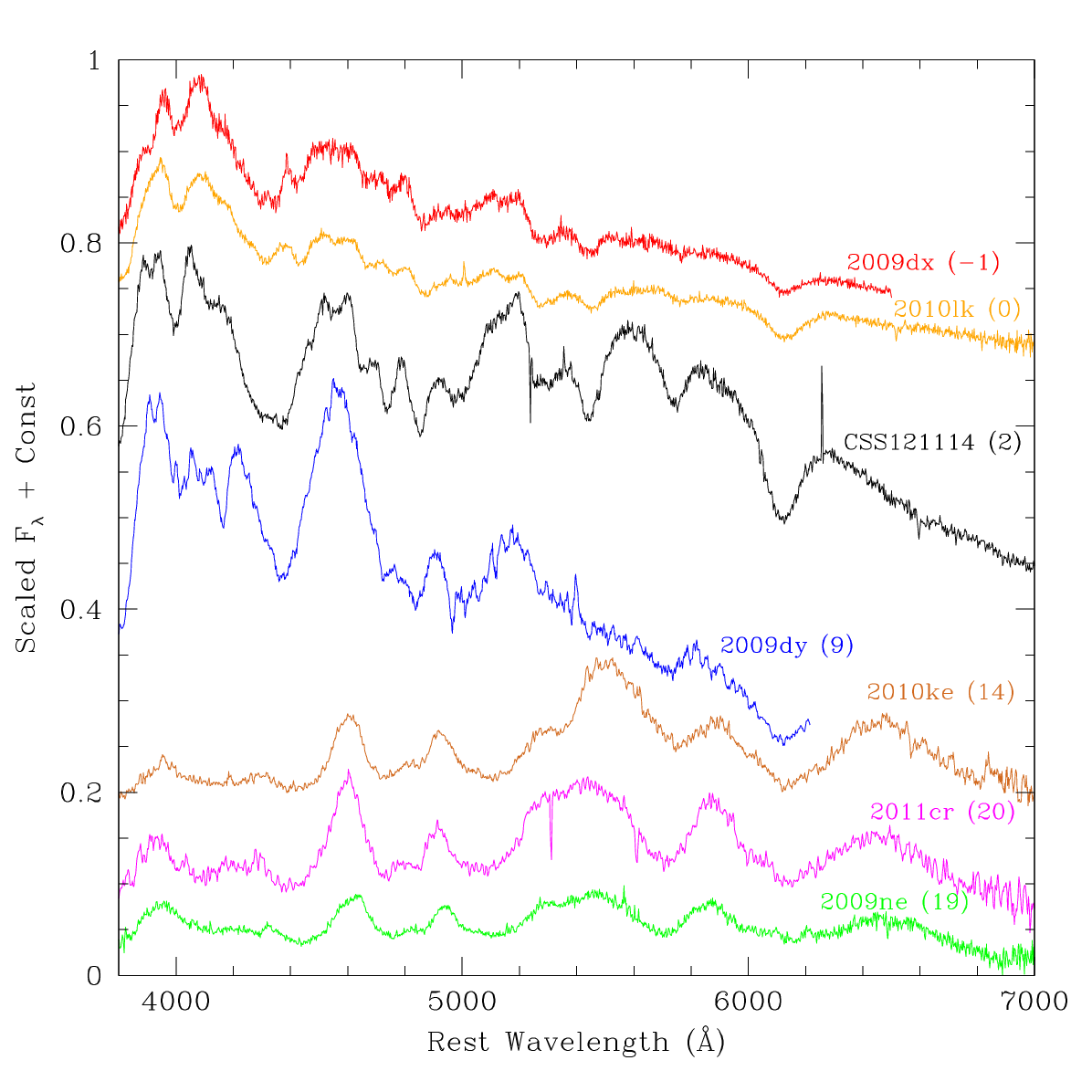}
\caption{\label{SpecIa}
Example spectra for nearby type-Ia supernovae discovered by CRTS.
The age of events in days (in parentheses) relative to maximum light are marked and sorted 
from youngest (top) to oldest (bottom).
}
}
\end{figure}

SN-Ia are the most common class of supernova. Thus, even though we did not
target such events, almost half of the spectra that we obtained were type-Ia's. 
These events are of particular interest in cosmology due to their use as standard 
candles. The low redshift supernovae anchor high-redshift SN-Ia samples for dark 
energy, trace large-scale structure, and help measuring $H_0$ (Dhawan et al.~2022). 
However, the use of SN-Ia for these purposes generally requires well-characterized 
lightcurves in standard passbands. 
Thus, rather than attempting to perform the dedicated follow-up required to use 
the SN-Ia as cosmological probes, we announced all of our SN-Ia classifications 
publicly as quickly as possible to aid such efforts.

In Figure \ref{SpecIa}, we present spectra for a sample of SN-Ia that we spectroscopically
confirmed. All of these events were publicly announced in ATels or CBATS. Out of the 321
classified spectra, $\sim$43\% were of normal type-Ia supernovae. 
Other similar-sized samples have been released by nearby supernovae surveys dedicated 
to type-Ia research, including those presented by Copin et al.~(2009)\nocite{Cop09} and 
Burns et al.~(2018)\nocite{Bur18}. However, much larger samples of local type-Ia's have 
recently been presented, e.g. Silverman et al.~(2012)\nocite{Sil12} and
Dhawan et al.~(2022)\nocite{Dha22}. Therefore, we have not conducted a detailed
analysis of these events.

\subsection{Type-II Supernovae}

Type-II supernovae are due to core collapse (CC) events and are usually easily identified
by broad hydrogen emission lines present within their spectra. As with the type-Ia supernovae
we did not specifically target common type-II's (such as type IIP's and IIL's). Nevertheless,
almost half of the supernovae we observed within two weeks of detection were type-II's.

\begin{figure}{
\includegraphics[width=84mm]{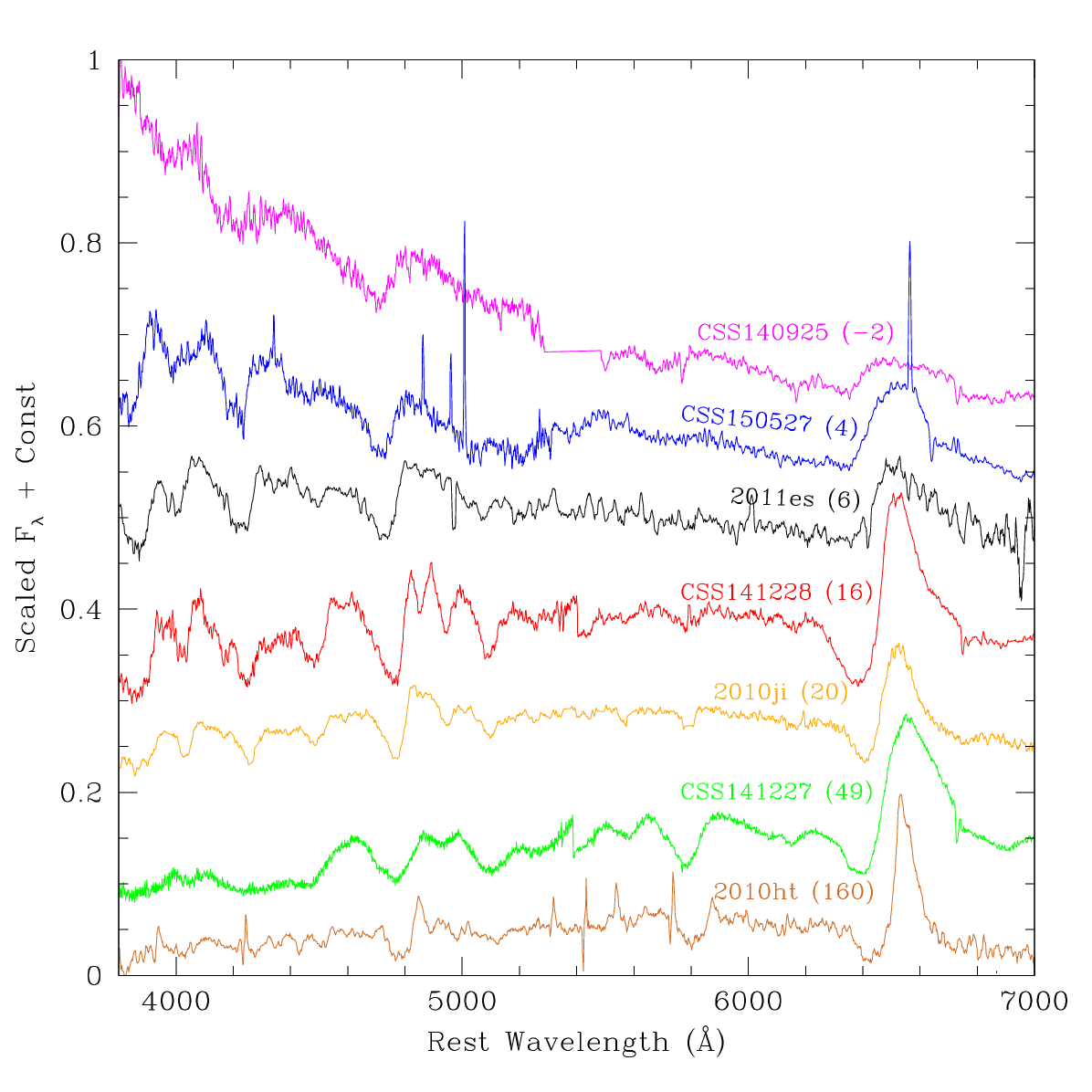}
\caption{\label{SpecIIP}
As in Figure 1, but for a samples of type-IIP supernovae spanning a range 
of ages, sorted vertically from youngest to oldest.
}
}
\end{figure}

Of the 122 type-II supernovae we observed, 38 were classified as type-IIP. In Figure 2, we present
examples of the type-IIP spectra we obtained. One of the most unusual type-II's discovered 
was supernova CSS141118:092034+504148. Based on our public discovery information, the event 
was spectroscopically confirmed as a type-IIP by Li et al.~(2015)\nocite{LiW15}. 
The event was much later noted by Arcavi et al.~(2017)\nocite{Arc17} to have had an earlier
detection within the private data obtained by PTF.
The authors suggested it was an unusually long-lived type-IIP supernova. Recently, Wang et al.~(2022)\nocite{Wan22} 
have suggested that the event was a pulsational pair-instability supernova. However, had 
the early PTF detection been made public in 2014, stronger constraints would likely have 
been placed on this current classification.

\subsection{Superluminous Supernovae}

\begin{figure*}{
\includegraphics[width=85mm]{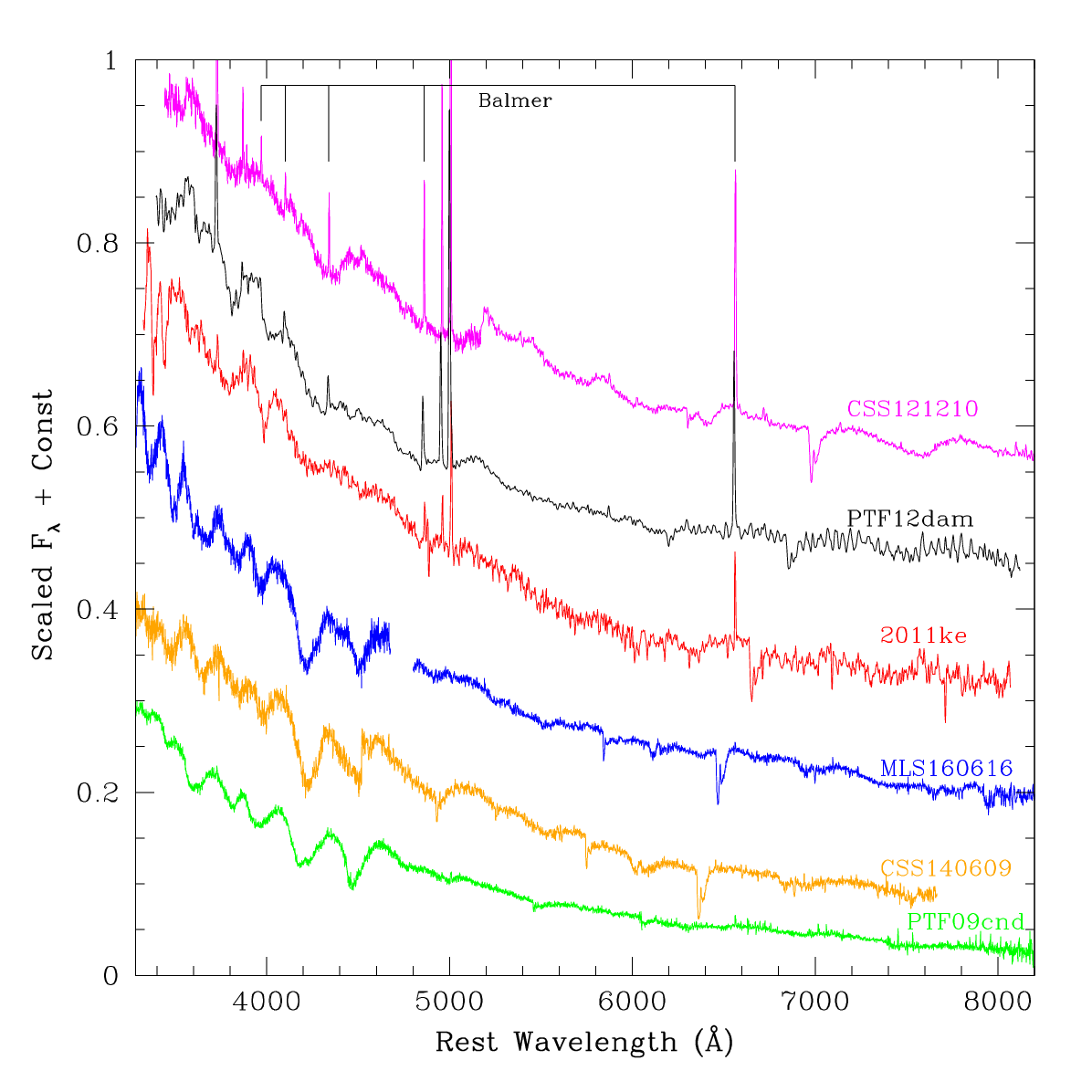}
\includegraphics[width=85mm]{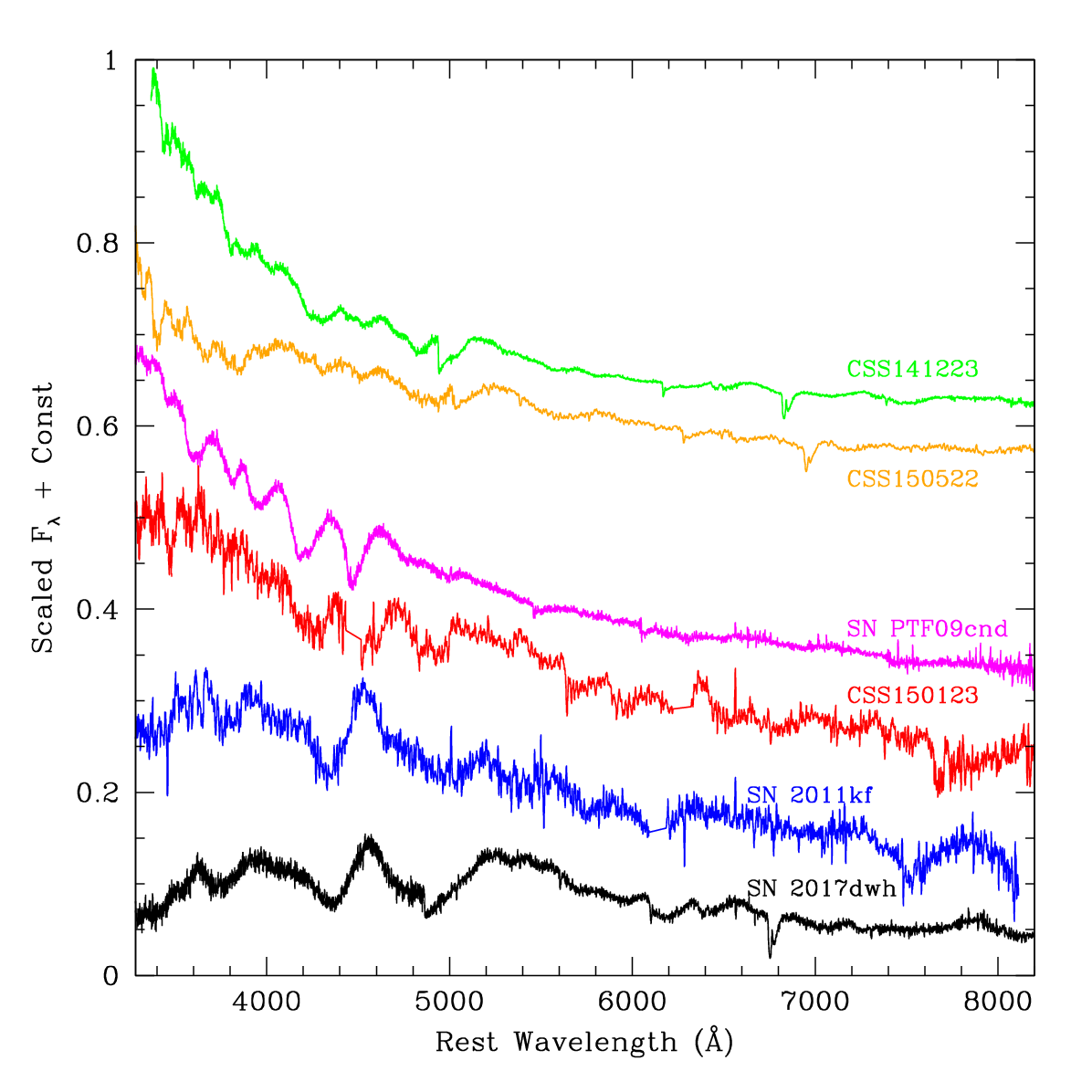}
\caption{\label{SpecSL}
Spectra of SLSN-I discovered by CRTS compared to known PTF SLSN-I PTF09cnd and PTF12dam.
In the left panel, from top to bottom we present CRTS CSS121210 (= CSS121210:112615+144039),
SN 2011ke (= CSS110406:135058+261642), MLS160616 (= MLS160616:160420+392813), and
CSS140609 (= CSS140609:224011+095209). In the right panel, we present CRTS SLSN-I
CSS141223 (= CSS141223:113342+004332, 2015bn), CSS150522 (= CSS150522:180139+381750)
CSS150123 (= CSS150123:102121+524752), SN 2011kf (= CSS111230:143658+163057) and
SN 2017dwh (= CSS170425:143443+312917).
}
}
\end{figure*}

Among the supernovae detected and spectroscopically confirmed by CRTS, 15 are type-I superluminous 
supernova (SLSN-I) candidates. However, the classification of six of these events still remains unclear. 
In Figure \ref{SpecSL}, we present the spectra of SLSN-I discovered by CRTS, compared to the known SLSN-I's
PTF09cnd and PTF12dam (Quimby et al.~2011\nocite{Qui11}, 2012\nocite{Qui12}). Additional SLSN-I discovered
by CRTS and discussed elsewhere included CSS100313:112547-084941 (Mahabal et al.~2010\nocite{Mah10}; designated
SN 2010gx; alias PTF10cwr, Quimby et al.~2010\nocite{Qui10}), CSS120121:094613+195028 (designated SN 2012il, 
Drake et al.~2012a\nocite{Dra12a},b\nocite{Dra12b}; alias PS1-12fo, Smartt et al.~2012\nocite{Sma12}), and 
CSS090802:144910+292510 (designated SN 2009jh, Drake et al.~2009b\nocite{Dra09b}; alias PTF09cwl, Quimby 2009\nocite{Qui09}).

In addition to these events, SLSN-I $\rm CSS141223:113342+004332$ (CSS141223) was discovered by CRTS and
classified as a very likely supernova candidate on 2014 December 23. Approximately two months later the
event was redetected by PanSTARRS-1 (PS1) and named PS15ae \cite{Nic16}. The event was classified as an
SLSN-I by the Public ESO Spectroscopoc Survey for Transient Objects (PESSTO, Le Guillou et al.~2015\nocite{LeG15}),
and although it was discovered in 2014, it was designated SN 2015bn. Our confirmation spectrum of
CSS141223 \cite{Dra15} is presented in Figure \ref{SpecSL} and is consistent with the PESSTO classification.
However, the spectral features are different than the other SLSN-I that we discovered.

\subsection{Interacting Supernovae}

In the late stages of their evolution, stars can expel material from their outer envelopes. When these stars
ultimately become supernovae, the shells of expanding material are illuminated by the radiation released during
the event. These shells of circumstellar material (CSM) typically contain hydrogen and helium ejected from the
outermost layers. 

Type-IIn supernovae are well-known CC supernovae that exhibit Balmer emission due to CSM. Recently, many other
types of supernovae have been discovered that exhibit CSM interactions. For example, type-Ia supernovae are well
known for the lack of hydrogen and helium in their spectra \cite{Fil97}. These events are believed to be due to
the thermonuclear explosions of accreting CO white dwarfs (Iben \& Tutukov 1984\nocite{Ibe84}; Webbink 1984\nocite{Web84}).
However, moderately large numbers of type-Ia's that exhibit hydrogen within their emission spectra (SN Ia-CSM) 
have been discovered (e.g. Sharma et al.~2023\nocite{Sha23}). 

Evidence for interacting shells of material has also been observed in a small fraction of stripped envelope CC
supernovae (type Ib and Ic) that lack broad absorption lines due to hydrogen. These supernovae emit narrow emission
lines and are denoted as SN-Ibn (Foley et al.~2007\nocite{Fol07}, Pastorello et al.~2008\nocite{Pas08}) and
SN-Icn (Gal-Yam et al.~2022\nocite{Gal22}, Perley et al.~2022\nocite{Per22}).

\subsubsection{Type-IIn Supernovae}

The most common supernovae to exhibit CSM interactions are SN-IIn. These events exhibit a wide range of timescales,
peak luminosities and decline rates \cite{Tad15}.
Type-IIn are thought to result from massive stars (such as luminous blue variables, LBVs) that have 
lost their hydrogen envelopes through wind-driven mass-loss and outbursts (Gal-Yam et al.~2007\nocite{Gal07}, 
Smith et al.~2008\nocite{Smi08}). However, there is also evidence that some of these events are due to red 
supergiants \cite{Smi09}.

\begin{figure*}{
\includegraphics[width=85mm]{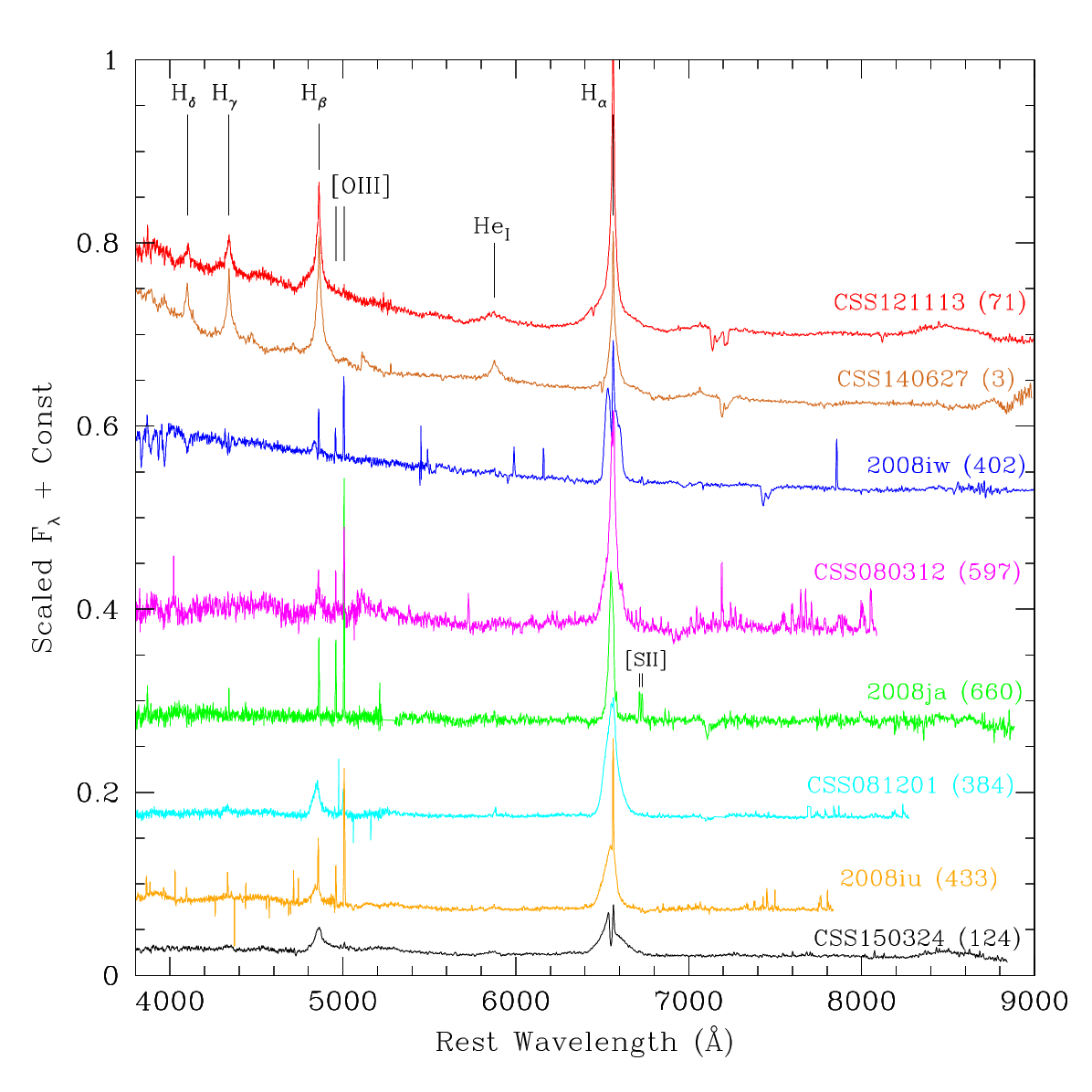}
\includegraphics[width=85mm]{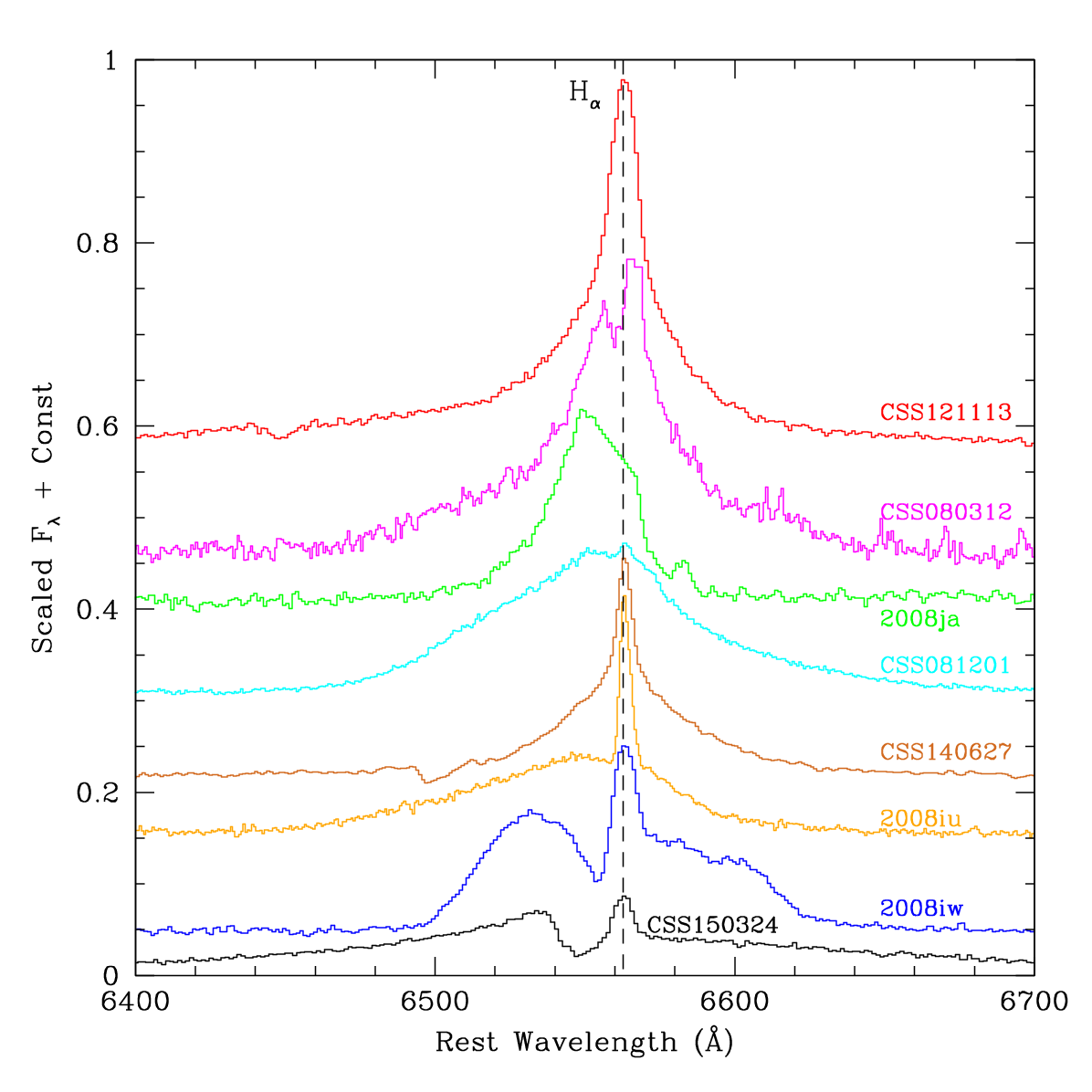}
\caption{\label{IIn} 
Spectra of type-IIn supernovae discovered by CRTS.
In the left panel, we show the spectra of type-IIn supernovae discovered by CRTS with 
best matching SNID {\protect\cite{Blo07}}. The spectra span post-maximum ages (in parentheses) 
rangeing from 3 to 660 days. In the right panel, we show the diversity of $\rm H_{\alpha}$ 
line profiles for these same events.
}
}
\end{figure*}

In Figure \ref{IIn}, we present the spectra of eight SN-IIn discovered by CRTS along with a zoomed inset
of the $\rm H_\alpha$ regions of these events. The CRTS spectroscopic follow-up was particularly sensitive
to long-timescale type-IIn, with 63 of the extragalactic transients having this classification. The wide
range of Balmer line widths and shapes shown in Figure \ref{IIn} is likely due to the diversity of the
distribution of material surrounding the progenitor stars as well as the anisotropy of the energy emitted.

\subsubsection{SN 2008iy}

\begin{figure*}{
\includegraphics[width=85mm]{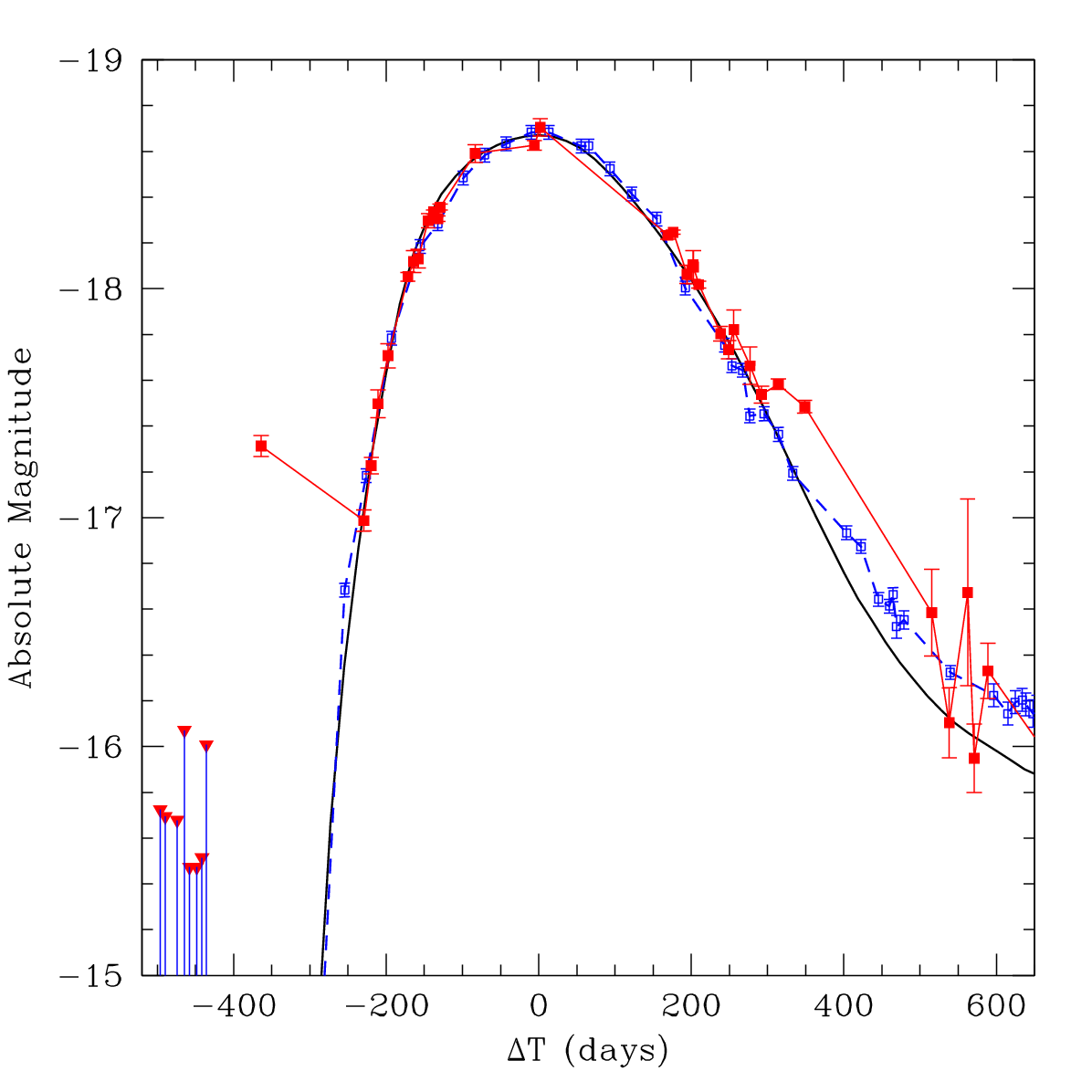}
\includegraphics[width=85mm]{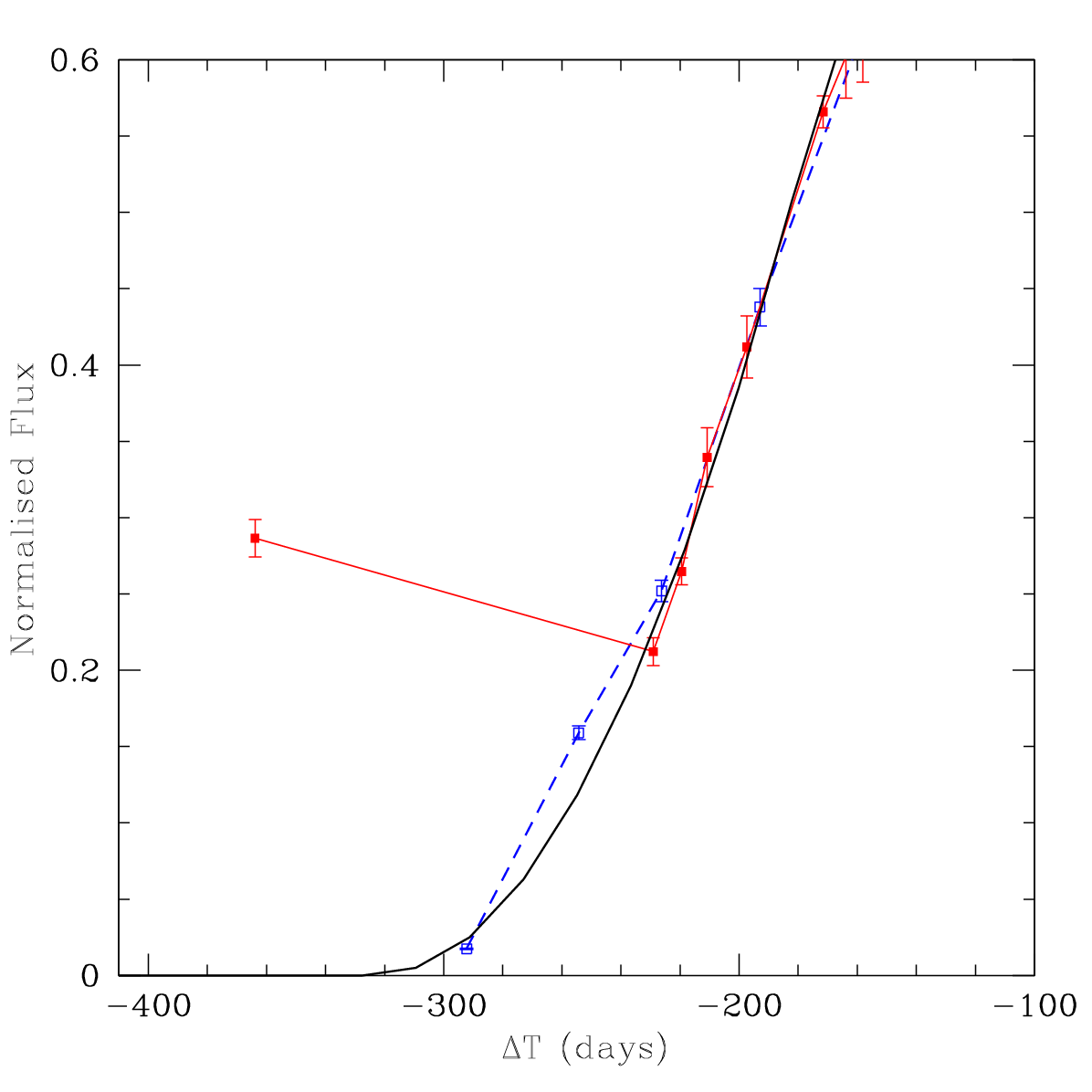}
\caption{\label{LCIY} 
The nightly-combined rest-frame lightcurves of SN 2008iy (red, solid line) compared with other 
supernova light curves matched to peak magnitude and timescale (using stretch factor $S$). 
Left panel: the observed absolute magnitude light curve for SN 2008iy (red points;
triangles upper limits, squares detections); type-IIn supernova SN 2006gy {\protect\cite{Smi07}} 
scaled with $S=4.8$ and matched to 2008iy in peak brightness (blue squares and dashed-line); empirical type-Ia 
supernova model Parab-18 (black line, Goldhaber et al.~2001{\protect\nocite{Gol01}}) scaled with $S=18.2$ and matched
in peak brightness. Right panel: normalised fluxes for the same set of supernova light curves and stretches factors
as left panel.
}
}
\end{figure*}

\begin{figure}{
\includegraphics[width=85mm]{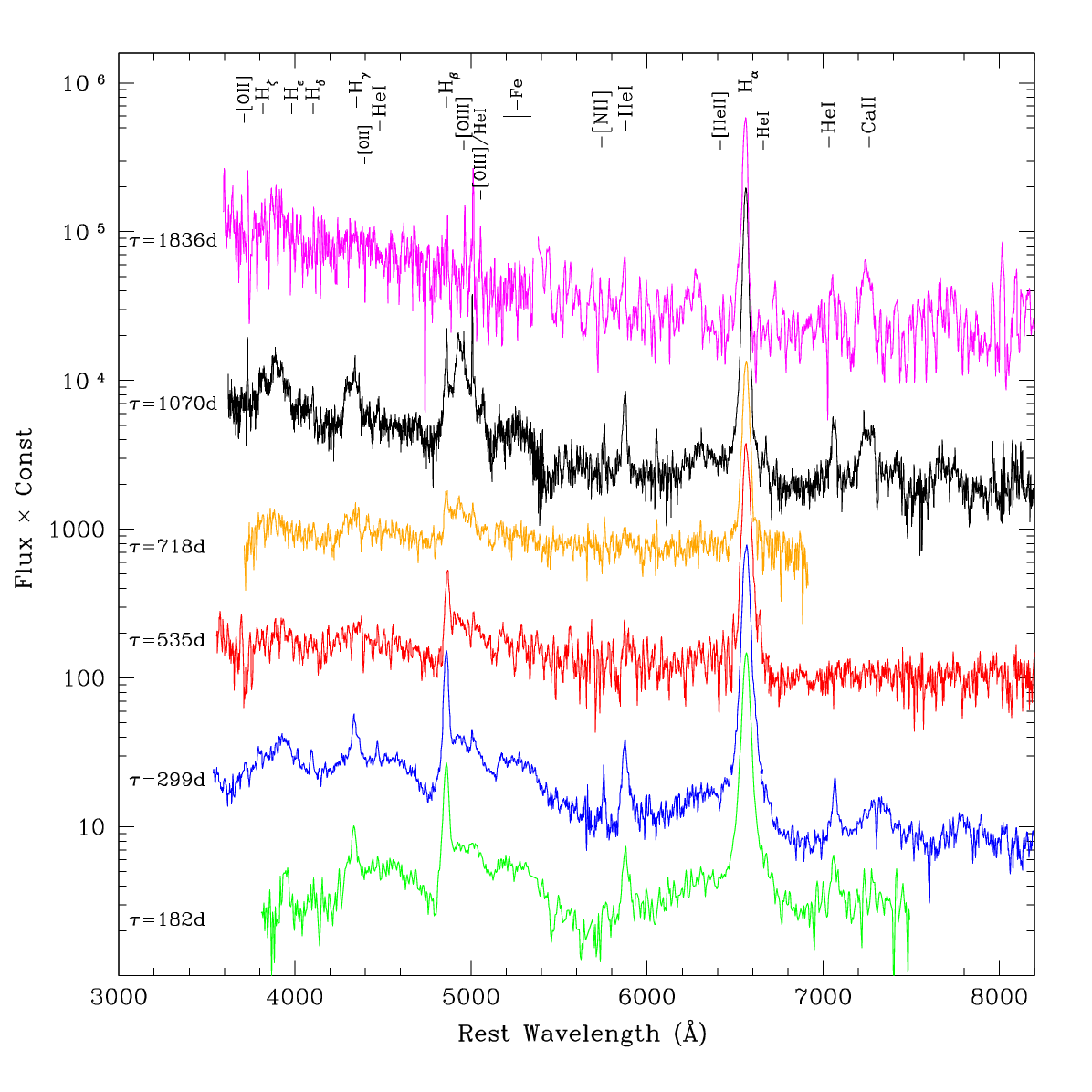}
\caption{\label{SpecIY}
Spectra of type-IIn SN 2008iy obtained between 182 and 1836 days past maximum 
light with the SMARTS 1.5m + RCSPEC, Palomar 5m + DBSP, and Keck + LRIS.
}
}
\end{figure}

One of the type-IIn supernovae that we followed the most closely was SN 2008iy. This event was
discovered by CRTS on 2008 September 28 \cite{Dra08b}. Given the exceptionally long rise-time of
the event, we announced the discovery via ATels and began multi-colour photometric follow-up with 
the Palomar 1.5m the following day. We obtained spectroscopic observations of the transient with
the SMARTS 1.5m telescope \cite{Mah09} 
and found the transient to be an SN-IIn well past maximum, with broad hydrogen and helium
emission lines, and a significant hump near 510 nm due to a mixture of iron lines.

Miller et al.~(2010)\nocite{Mil10} analysed the rise of the event based on an incomplete set
of raw CRTS photometry and suggested that the event had undergone a 400-day rise. To determine
a better estimate of the true timescale of the event and remove errors due to blending with a
nearby bright star, we created a high signal-to-noise ratio template image and performed image 
subtraction \cite{Tom96} 
on the complete set of CSS images. We then extract photometry for each of the images and combine
the set of four measurements from each night to create a light curve for the transient. 

The combined nightly lightcurve and the individual observations show that SN 2008iy faded significantly 
following an initial outburst. Therefore, the event likely did not rise for the full 400 days as previously 
suggested. Instead, the progenitor underwent an extremely luminous outburst ($\rm M_{V} = -17.4$) that began 
$> 60$ days before the supernova explosion. Additional photometry also suggests that an earlier outburst may 
have occurred in April 2004. This kind of pre-supernova brightening was also observed in type-IIn SN-2009ip. 
The outbursts of SN-2009ip occurred as it evolved from an LBV to a true type-IIn supernova \cite{Smi22}. 
However, in the case of SN-2009ip, the pre-supernova outbursts only reached $\rm M_{V} \sim -14.8$.

In Figure \ref{LCIY}, we show the absolute and normalized rest-frame flux lightcurves 
of SN 2008iy compared with other supernova lightcurves and models that have been stretched 
in time (by a factor $S$) to match. In this comparison, we include energetic type-IIn 
supernova SN 2006gy \cite{Smi07} 
with $S = 4.8$, an empirical supernova model SCP1997 with $S = 18.2$ (Goldhaber et al.~2001), 
and SN 2009ip (Prieto et al.~2013\nocite{Pri13}) with $S = 17.9$.
To illustrate how poorly the initial photometric measurements match typical supernova rising
phases, we include a zoom-in plot of the initial outburst. Fitting of the CSS lightcurve reveals that 
maximum light occurred on $\rm MJD=54735 \pm 5$ with a peak absolute magnitude of $\rm M_{V} = -18.6 \pm 0.1$.

To identify the host galaxy we coadded SDSS $u$,$g$,$r$, $i$ and $z$ images taken at the location of SN 2008iy
seven years prior to discovery. 
We identify the host galaxy as a source with $r \sim 23$ that lies within one arcsecond of the supernova. 
Based on the supernova's redshift ($z = 0.041$), the host is found to have $\rm M_r \sim -13$, making it 
a dwarf galaxy that is $\sim 100$ times fainter than the Large Magellanic Cloud \cite{Mah09}. 

In Figure \ref{SpecIY}, we plot spectra of SN 2008iy taken over five years with the SMARTS-1.5m + RCSPEC,
Palomar 5m + DBSP, 
and Keck + LRIS. 
Interestingly, we see that the spectra of the supernova show very little evolution in the first year with
He-I and Ca-II varying in strength more than a year after maximum. After five years, intermediate-width
$\rm H_\alpha$ ($3.5 \,\rm km\, {\rm s}^{-1}$) lines remain clear, while other Balmer lines have faded
completely. The $\rm H_{\alpha}$ emission remains and evolves slowly from the confirmation spectrum
to $> 1800$ days after peak brightness.

\subsubsection{Interacting Type-I Supernovae}

As noted above, SN Ia-CSM are a rare subtype of type-Ia that initially appear as normal Ia's yet exhibit
strong Balmer lines as they evolve. These supernovae provide constraints on which of the two main type-Ia
progenitor scenarios is the most likely. In the double-degenerate (DD) scenario, SN-Ia arise when two CO white
dwarfs merge (Iben \& Tutukov 1984\nocite{Ibe84}; Webbink 1984\nocite{Web84}), while in the single-degenerate
(SD) scenario, the explosion is caused by accretion onto the WD by a companion \cite{Whe73}. There is significant
evidence for a DD-based origin based on individual events such as SN 2011fe (e.g. Bloom et al.~2011\nocite{Blo11},
Nugent et al.~2011\nocite{Nug11}, Brown et al.~2012\nocite{Bro12}). However, the occurrence of SN Ia-CSM is
more consistent with the SD scenario as SN Ia-CSM spectra and energetics appear more consistent with the
thermonuclear explosions expected for Ia's, rather than core-collapse events (Aldering et al.~2006\nocite{Ald06}, 
Prieto et al.~2007\nocite{Pri07}).

\begin{figure*}{
\includegraphics[width=85mm]{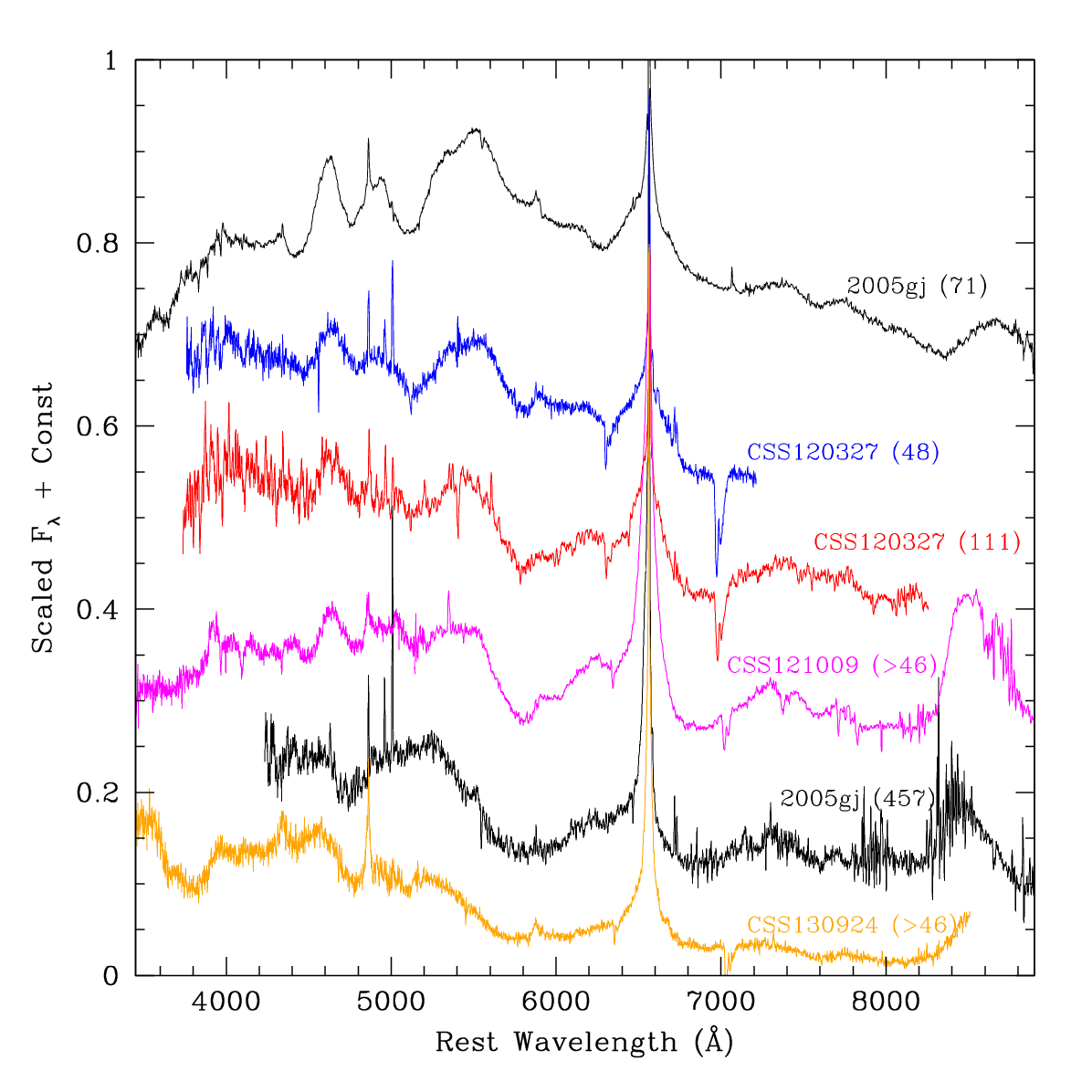}
\includegraphics[width=85mm]{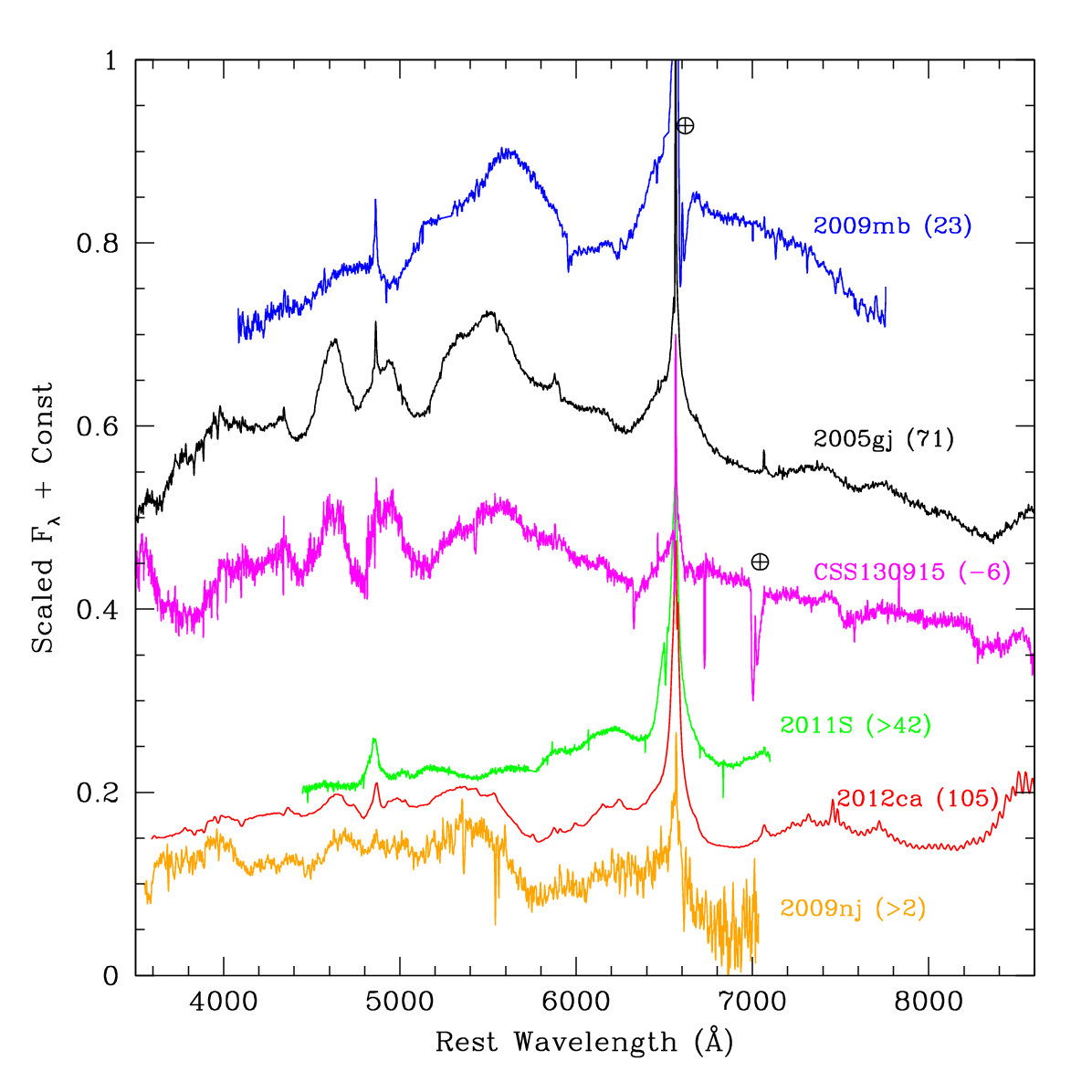}
\caption{\label{IaCSM} 
Spectra of type-Ia-CSM supernovae discovered by CRTS compared to spectra
of previously known Ia-CSM SN 2005gj and SN 2012ca.
In the left panel, we present the spectra of 
CSS120327 (= CSS120327:110520-015205), 
CSS121009 (= CSS121009:025917-141610), 
and CSS130924 (= CSS130924:080531+321902).
In the right panel, we present the spectra of 
SN 2009mb (= CSS091018:091109+195945) 
CSS130915 (= CSS130915:231108+173855),
SN 2011S (= CSS110114:091354-170039), 
SN 2009nj (= CSS091018:091109+195945). 
}
}
\end{figure*}

In total, we obtained 23 spectra for 15 CRTS supernovae that are most consistent with type Ia-CSM. However, the
classification of six of these candidates is unclear due to their similarity with type-IIn spectra (Leloudas et al.~2015\nocite{Lel15},
Sharma et al.~2023\nocite{Sha23}). In Figure \ref{IaCSM}, we present a sample of likely SN Ia-CSM candidates
discovered by CRTS in comparison to the known type Ia-CSM SN 2005gj (Aldering et al.~2006\nocite{Ald06},
Prieto et al.~2007\nocite{Pri07}) and SN 2012ca (Drescher et al.~2012\nocite{Dre12}, Inserria et al.~2016\nocite{Ins16}).
Analysis based on spectroscopic follow-up of four additional CRTS Ia-CSM candidates 
(CSS080505:155415+105825=SN 2008cg, Drake et al.~2008a\nocite{Dra08a}, Blondin 2008\nocite{Blo08}; 
CSS101008:001049+141039, Drake et al.~2010a\nocite{Dra10a}, Arcavi et al.~2010\nocite{Arc10}; 
CSS111128:113705+152814, SN 2011jb, Drake et al.~2011c\nocite{Dra11c},2011d\nocite{Dra11d}; 
CSS120327:110520-015205, Drake et al.~2012b\nocite{Dra12b}) 
was presented in Silverman et al.~(2013)\nocite{Sil13} and a detailed analysis of SN Ia-CSM event 
CSS130403:150213+103846 (= SN 2013dn, Drake et al.~2013d\nocite{Dra13d}) was presented in Fox et al.~(2015)\nocite{Fox15}.

\subsection{Stripped-envelope Supernovae}

Stripped-envelope supernovae (SESNe) include many types of supernovae that lack hydrogen in their spectra. 
Type-Ic are deficient in helium and hydrogen, while type-Ib's are deficient in hydrogen. The progenitors 
of these events lose their envelopes via stellar winds, binary interactions and eruptions \cite{Smi14}, 
and are thought to result from Wolf-Rayert stars \cite{Woo95}. Interacting supernovae without signs of
hydrogen are rare. Type-Ibn's have He-rich CSM, and type-Icn shows signs of a stripped C-O layer with H/He-poor CSM.
The supernova SN 2021ocs is a rare example where C and He lines are absent, and the progenitor star's O-Mg
layer was exposed \cite{Kun22}. Such supernovae can exhibit bumps or secondary peaks in their lightcurves.

Examples of supernovae with such secondary peaks include SLSN-Ic LSQ14bdq, which faded from an initial peak
for 10 days before the event brightened to its maximum \cite{Nic15}. 
SLSN-I 2017egm also had a slight second bump ($\sim 180$ days after peak) that was $\sim$2.5 
mags fainter than the initial maximum \cite{Zhu23}. Normal type-Ic SN iPTF15dtg \cite{Tad16} 
exhibited slight bumps at both early and late times. Similarly, the broadline (BL) type-Ic SN 2020bvc 
exhibited a bump at early times along with radio and X-rays \cite{Ho20}. 
Lastly, recent supernova SN 2022xxf had a secondary peak \cite{Kun23}. This supernova was a
type-Ic/BL that rose to a similar absolute magnitude as the peak but did so within 80 days. The
latest spectrum of this object, taken at 122 days past maximum, shows strong Ca and O rather
than narrow Fe and Mg and is similar to the spectrum of 
SN 2012au. The event shows a gradual rise of $< 1.5$ magnitudes over $\sim 40$ days.

\subsubsection{Extreme Rebrightening Supernova CSS111225}

\begin{figure}{
\includegraphics[width=85mm]{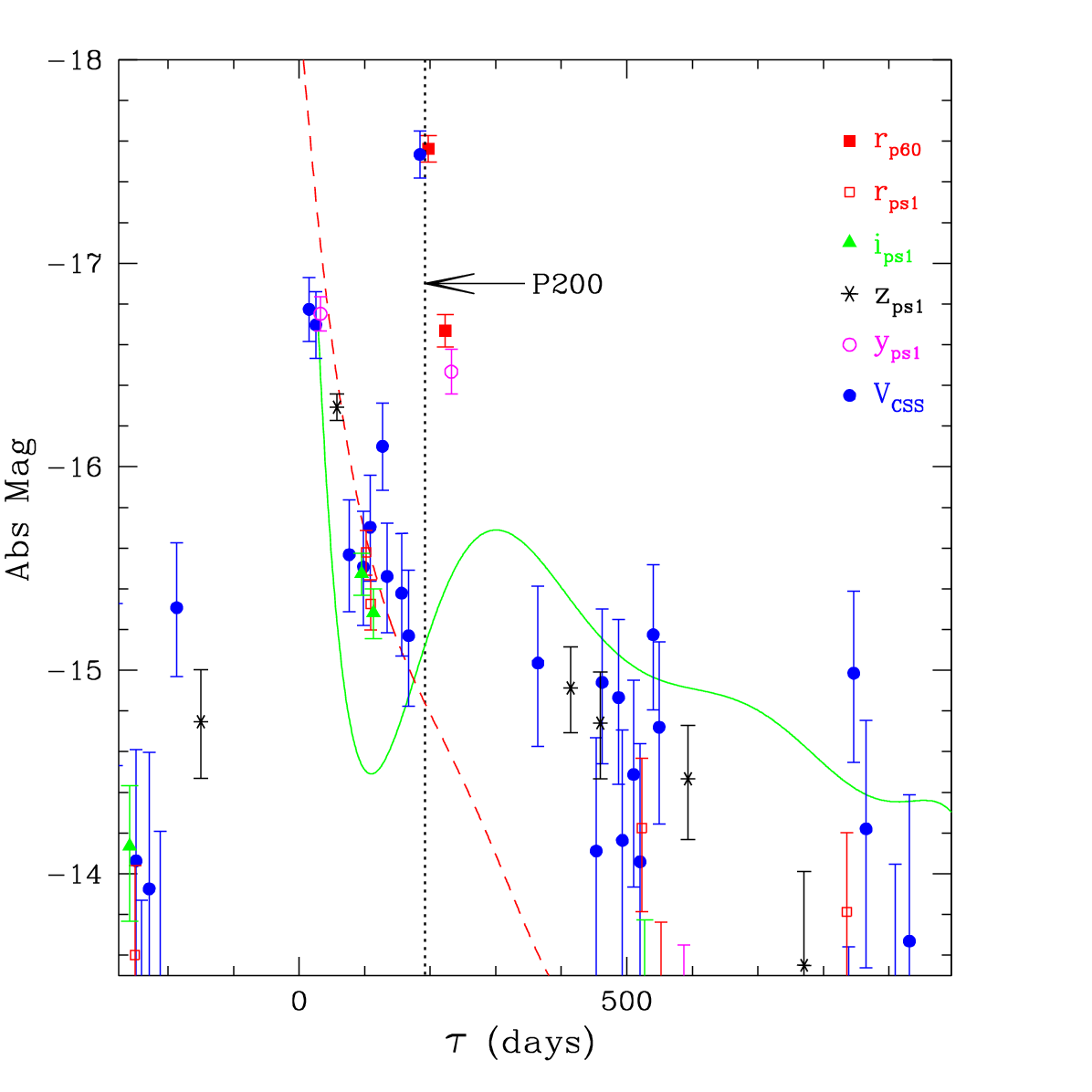}
\caption{\label{LCCSS111225}
The multi-wavelength lightcurve of type-In supernova CSS111225 compared to polynomial fits to the
lightcurve of SN 2012au (red-dashed line, Pandy et al.~2021{\protect\nocite{Pan21}}) and public 
ZTF $r$-band photometry for SN 2019oys (green line, Sollerman et al.~2020{\protect\nocite{Sol20}}). 
The vertical dotted line shows the epoch of the spectroscopic follow-up taken with P200 and DBSP.
Each CRTS data point is the nightly average of four observations.
}
}
\end{figure}

Supernova CSS111225:140122+161705 (hereafter CSS111225) was discovered on 2011 December 25 at $V$=18.7,
and was reported via CBAT TOCP as a possible supernova (PSN J14012159+1617047). The host galaxy is 
visible in SDSS images with a brightness of $r=19.4$. The combined nightly CRTS photometry shows
that the transient declined by two magnitudes over 150 days. However, the object re-brightened by
more than two magnitudes between 2012 May 25 and June 11. As there were no known examples of
supernovae with this behaviour at the time, we carried out follow-up spectroscopy and photometry.
The P60 $r$-band follow-up photometry confirmed the rebrightening event. Subsequently, serendipitous
photometry of the event (in $r$, $i$, $z$ and $y$-bands), was released by PanSTARRS in DR2 \cite{Fle20}.
The PS1 observations confirm the initial event and show that the secondary peak declined for
at least 50 days.

In Figure \ref{LCCSS111225}, we present the multi-band lightcurve of CSS111225 including photometry
from CRTS, PS1 and P60. Compared to other supernovae, this event is exceptional since the secondary
peak is far brighter than the initial one, yet it occurred more than 150 days later. The absolute
magnitude of the second peak is far brighter than observed in the other supernovae having double
peaks. 

\begin{figure*}{
\includegraphics[width=170mm]{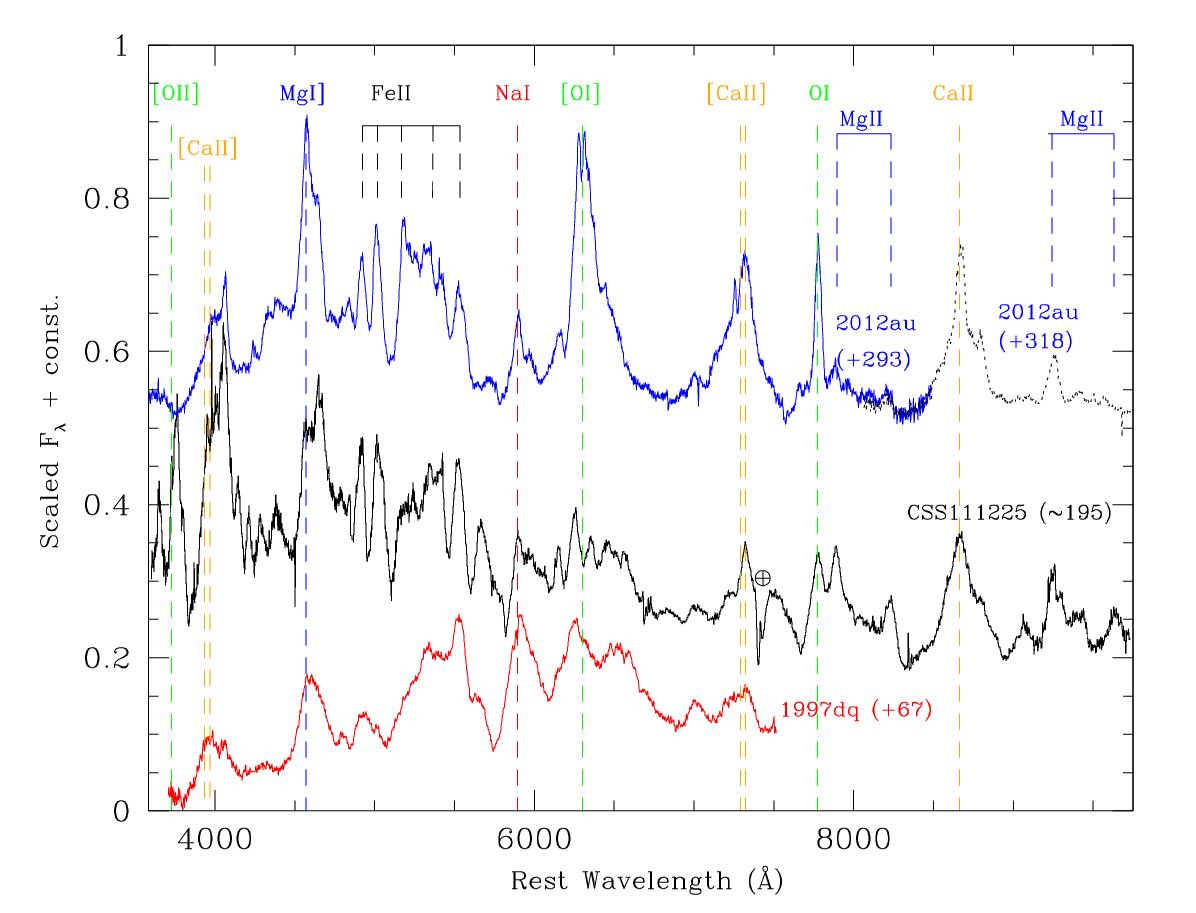}
\caption{\label{CSS111225}
The late-time spectrum of CRTS type-Ibn supernova CSS111225 compared to that
of broad-line type-Ic supernova SN 1997dq {\protect\cite{Maz04}}, 
and a composite spectrum for type-Ib SN 2012au {\protect\cite{Mil13}}. 
The combined SN 2012au spectrum includes data taken with the Multi-Mirror Telescope (MMT) at
293 days (solid blue line) past maximum light and with that from the Magellan Telescope data
taken 318 days (dashed black line). The SN 1997dq spectrum (red line) was taken with the
FLWO 1.2m telescope 67 days past maximum.
}
}
\end{figure*}

In Figure \ref{CSS111225}, we show the spectrum of CSSS11125, which exhibits 
Fe-O-Mg interaction but lacks hydrogen and helium. The optical spectrum is 
similar to the broad-line type-Ic SN 1997dq. We note that the slowly evolving 
SN 1997dq itself is unusual since it resembles the hypernova SN 1998bw \cite{Maz04}. 

Beyond $\rm 6500$\r{A}, the spectrum of CSS111225 resembles the unusual type-Ib SN 2012au, 
an event that may be a link between SLSNe and normal Ic's \cite{Mil13}. 
However, this apparent nebular spectrum also exhibits strong $\rm Mg_{II}$ lines.
Furthermore, unlike the continuous decline observed for SN 2012au, 
this supernova underwent an extreme re-brightening event similar 
to type-Ib supernovae SN 2019tsf and SN 2019oys \cite{Sol20}. 
Furthermore, as with SN 2019tsf, the spectrum does not show evidence for hydrogan and helium lines (that are 
typically found in CSM interactions). Thus the re-brightening event seems unlikely 
to be powered by interactions of a wind with the outer envelope.

Compared to other SESNe events, the host galaxy has very low luminosity ($\rm M_{V} \sim -15.8$).
We performed astrometry using the PS1 detections of the event and found no significant 
offset from the core of the host galaxy. No further outbursts are present in the 
CSS photometry taken up until 2024.

\subsubsection{Interacting Type-I Supernova 2009ny}

On 2009 October 20 CRTS discovered CSS091020:221124-091244 (aka SN 2009ny, Christensen et al.~2009\nocite{Chr09})  
with $V \sim 18.3$, offset from a faint SDSS galaxy ($r \sim 21.5$). We obtained a Gemini South spectrum
of SN 2009ny with GMOS on 2009 November 8 and found the closest SNID \cite{Blo07} match to be type-Ib
supernova SN 1999di. However, a review of the spectrum suggests that the event is a better match to a type-Ibn.

\begin{figure}{
\includegraphics[width=85mm]{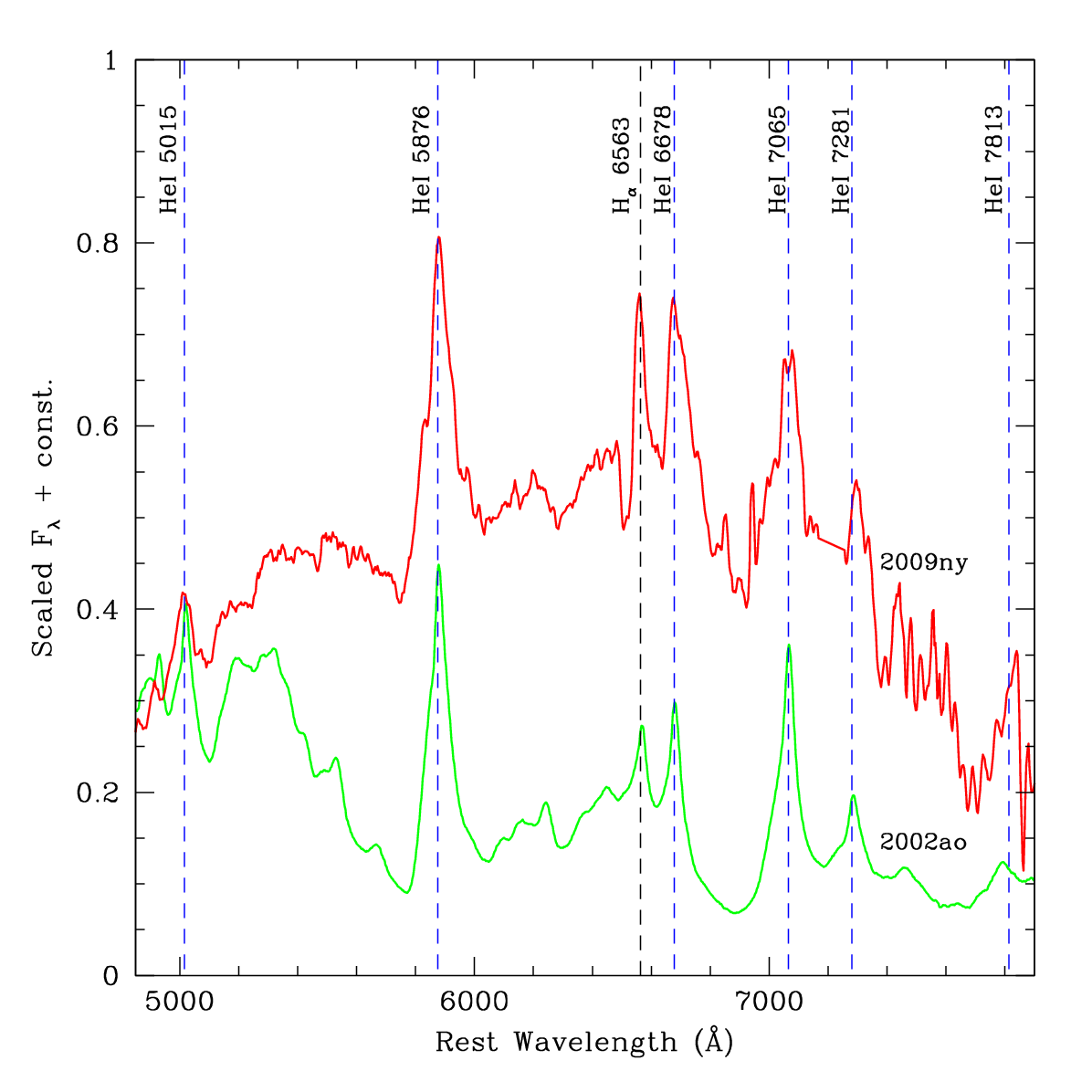}
\caption{\label{CSS091020} 
The spectrum of likely type-Ibn CRTS supernova (CSS091020:221124-091244)
SN 2009ny taken $\sim 23$ days after the peak, compared to the Keck LRIS spectrum 
of type-Ibn supernova SN 2002ao taken 20 days after discovery {\protect\cite{Fol07}},
obtained via the Weizmann Interactive Supernova Data Repository 
(WISeREP, Ofer \& Gal-Yam 2012{\protect\nocite{Ofe12}}).
}
}
\end{figure}

In Figure \ref{CSS091020}, we compare the spectrum of SN 2009ny to type-Ibn SN 2002ao. We find that the emission
lines are not as narrow as SN 2002ao's. The spectrum is similar to the day 45 spectrum of the radio-detected
type-Ia SN 2020ejy, which exhibits a helium-rich CSM \cite{Koo23}. 
However, SN 2009ny exhibits stronger $\rm H_{\alpha}$, with strong asymmetric peaks and P-Cygni features
on the blue wings and additional emission features on the red wings. In contrast, type-Ia He-CSM SN 2020ejy
exhibited attenuated red wings. Additionally, the dominant emission lines of SN 2002ao and SN 2009ny are
far broader than those of SN 2020ejy.

The lack of any early spectra for SN 2009ny means that we cannot definitively determine whether the event
originated as a type-Ia or type-Ib. Nevertheless, the match to type-Ibn SN 2002ao is much better than to
other types. The supernova itself rises to a peak of $\rm M_{V}= -19$ and declines in less than 50 days,
with no obvious plateau. This is also consistent with type-Ibn events (Pastorello et al.~2008\nocite{Pas08}).

\subsection{Host Galaxies}

To understand the nature of the galaxies hosting supernovae it is important to know the source distances. 
The SNID software \cite{Blo07} 
uses crosscorrelation to provide an estimate of the supernova type and the redshift. In most cases, this
software returns matches with uncertainties of $< 10\%$. However, in some cases, the spectra are seen to
contain narrow lines, such as [OIII] emission due to star formation, within the host galaxy. For these
events, we are usually able to obtain redshifts with uncertainties $< $1\%. Thus, by using host apparent
magnitudes corrected for foreground extinction, using reddening maps (Schlegel et al.~1998\nocite{Sch98},
Schlafly \& Finkbeiner 2011\nocite{Sch11}) and redshift-based distances, we can estimate the absolute magnitudes
for events with visible host galaxies. For the normal type-Ia supernovae in the sample we also crosschecked
that the spectroscopic distances were consistent with the peak brightness being $\rm M_{V} \sim -19.3$.

Although most of the supernova host galaxies in our survey are too faint to determine their Hubble types,
it can be informative to compare the luminosity function for the host galaxies that we discovered
with those from supernova surveys that concentrated on repeated observations of bright local galaxies.
For this reason we carried out a comparison with the Lick Observatory Supernova Search (LOSS, Leaman et al.~2011\nocite{Lea11}).

\subsubsection{Comparison with LOSS}

In contrast to CRTS, the LOSS survey was dedicated to the discovery of supernovae based on the repeated 
observations of a fixed set of 14,882 nearby ($z < 0.05$) galaxies. The LOSS galaxy sample was primarily 
selected from the well-known: Third Reference Catalogue of Bright Galaxies (RC3, de Vaucouleurs et al.~1991\nocite{deV91}). 
However, LOSS observations were taken with an average cadence of 9 days \cite{Lea11}, which is similar to CRTS.

\begin{figure*}{
\includegraphics[width=85mm]{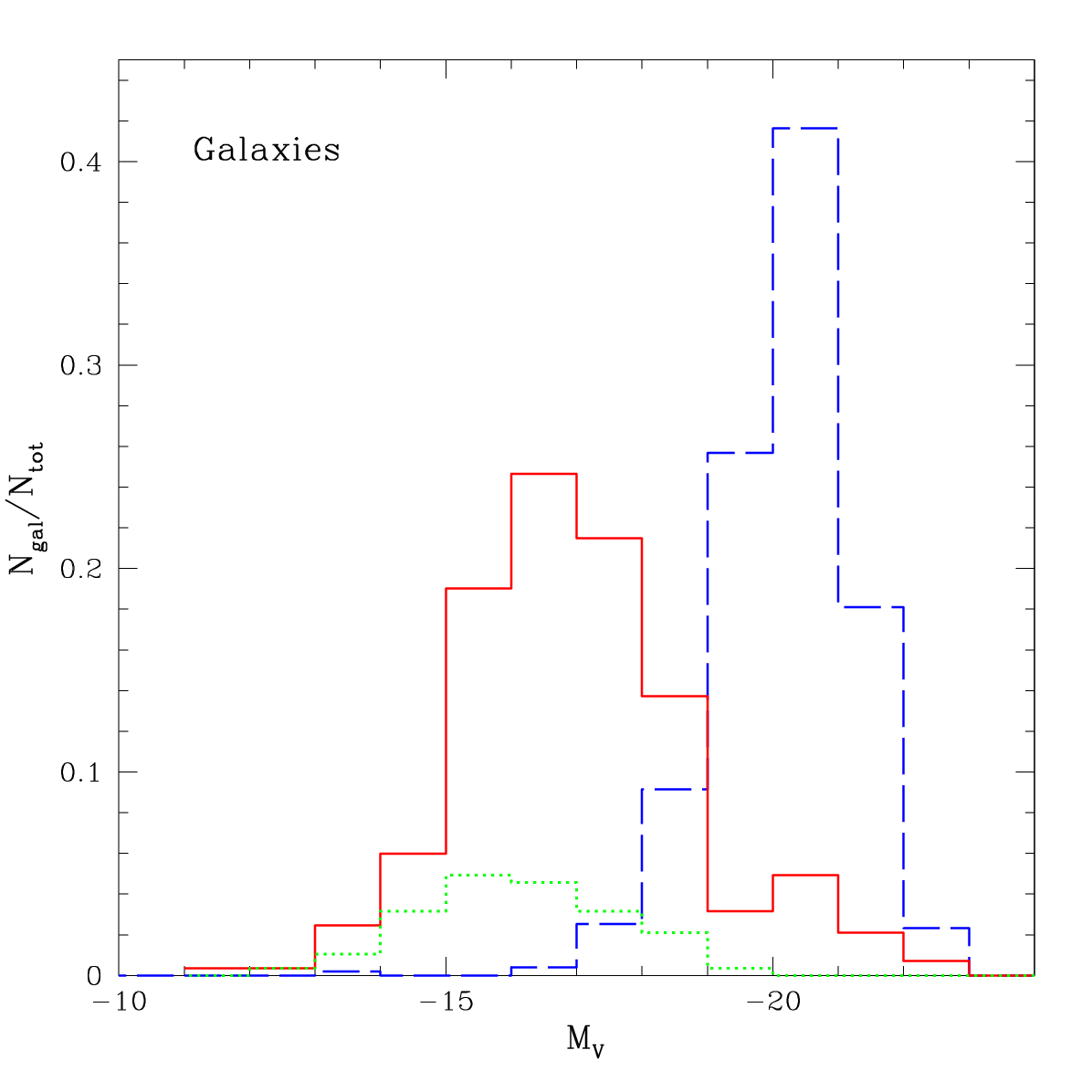}
\includegraphics[width=85mm]{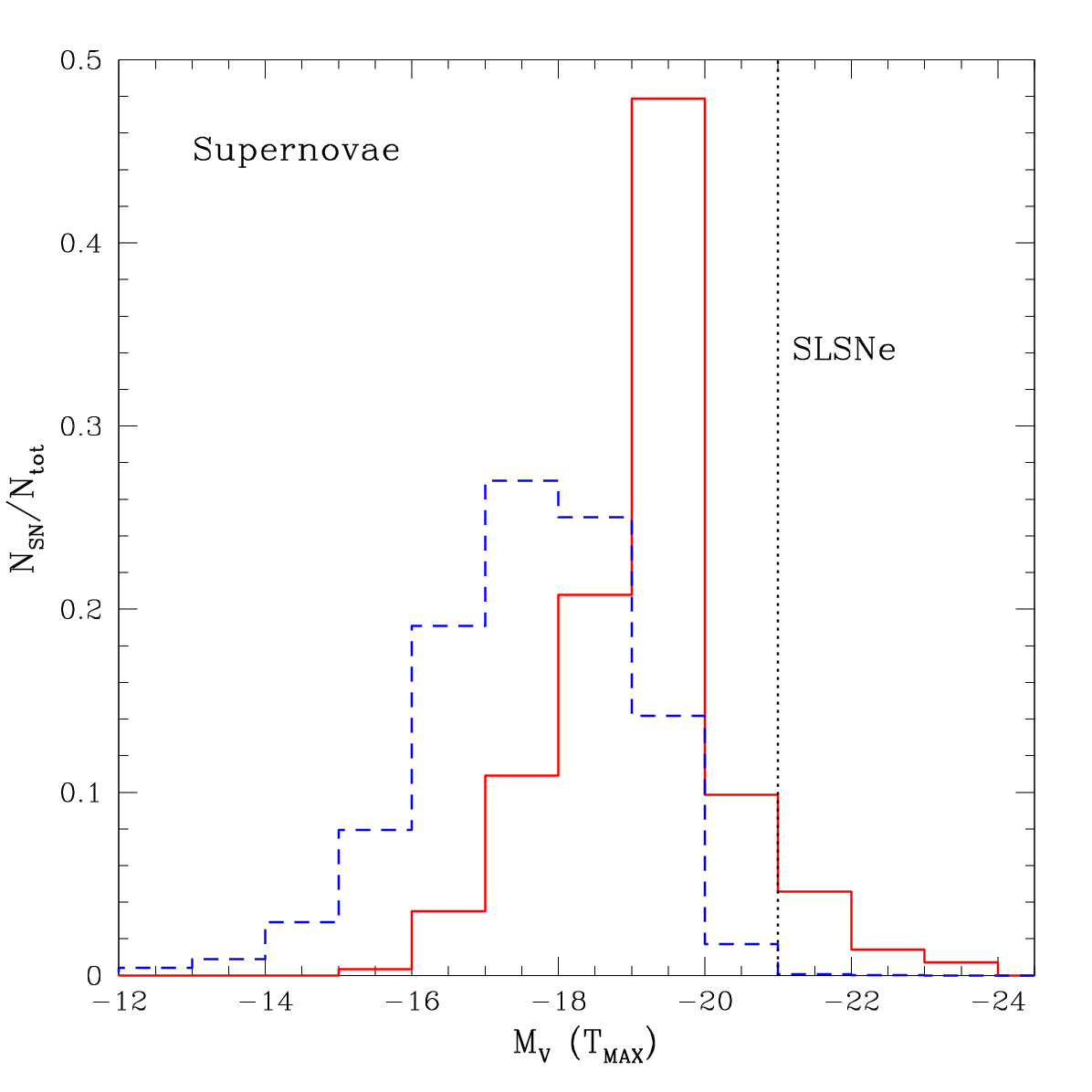}
\caption{\label{Hist}
Left panel shows the luminosity functions of supernova host galaxies.
The CRTS host galaxies are given by the solid red line, while the blue
dashed line shows the LOSS host galaxy sample. The green dotted line gives 
the upper limits in cases where no host galaxy was visible in CRTS.
In the right panel we show the distribution of supernova peak magnitudes.
The red line gives CRTS supernovae, while the dashed line gives 
LOSS supernovae. The vertical dotted line gives the division widely 
used to separate superluminous from normal supernovae.
}
}
\end{figure*}

In Figure \ref{Hist}, we show the distribution of supernova host galaxy luminosities and peak magnitudes
for CRTS discoveries compared to those found by LOSS. Here we see a stark difference in both the
distribution of host galaxies and supernova peak luminosities. The LOSS supernova sample lacks discoveries
in low-luminosity hosts, while the CRTS sample has few events in the brightest galaxies. The lack of LOSS
supernova discoveries in faint hosts is due to the clear selection bias. For example, thousands of
additional nearby galaxies were discovered by SDSS (e.g. Abazajian et al.~2009\nocite{Aba09})
and added to the HyperLeda catalogue \cite{Mak14} after the LOSS sample was defined. However, this
bias was inevitable given that the KAIT telescope used by LOSS had a FoV of $0.012\, \deg2$, and
thus limited to an effective survey area of $\sim 180\, \deg2$, while the CRTS survey
covered $\sim 20,\!000\, \deg2$. In contrast, since the CRTS project did not use image subtraction,
supernova within the brightest galaxies were missed.

A comparision between LOSS and PTF was undertaken by Graur et al.~(2017)\nocite{Grau17},
and revealed difference in the hosts populations for various types of supernovae.
As PTF undertook an untargeted survey for transient detection \cite{Rau09}, like the CRTS, the SN hosts 
distribution is expected to be similar. A more detailed set of demographics for SN host galaxies, 
based on a large sample of CC SN, has been undertaken for the PTF SN hosts by 
Schulze et al.~(2021)\nocite{Sch21}.

\subsubsection{Low-Luminosity Host Galaxies}

In Drake et al.~(2012a)\nocite{Dra12a} we showed that the distribution of supernova host magnitudes 
from CRTS peaks at $\rm M_{V}=-17.5$ with more than twenty spectroscopically confirmed 
supernova from the survey having hosts with intrinsic luminosities fainter than 
$\rm M_{V}=-15$ (e.g. Drake et al. 2008c\nocite{Dra08c}, 2009d\nocite{Dra09d}, 2009c\nocite{Dra09c}, 2011a\nocite{Dra11a}; 
Catelan et al.~2008\nocite{Cat08}).  
More recent CRTS supernovae continue this trend. For example, CSS140925:223344-062208 (SN 2015bs)
was classified by PESSTO \cite{Wal18} 
as a type II. Anderson et al.~(2018)\nocite{And18} determined that this event has a very faint
host with $\rm M_{r} = -12$.

Several projects have been carried out to search for so-called ``hostless supernovae''
(Gal-Yam et al.~2003\nocite{Gal03}, Sand et al.~2008\nocite{San08}).
These events were predicted to occur within intra-cluster material. Given that CRTS and other surveys
have discovered supernovae with host galaxies having $\rm M_{r} = -12$ and no clear cluster association,
it is clear that such surveys must go far deeper than the $\rm M_{R} \sim -15$ limit proposed by
McGee and Balogh~(2010)\nocite{McG10} and Zinn et al.~(2011)\nocite{Zin11} to select hostless supernovae.
Nevertheless, in the case of supernova MLS130531:100208+115705, we found evidence suggesting that supernovae
without clear hosts should occur due to the galaxy formation process. In this case, the type-Ia supernova
was found to likely be associated with a tidal stream orbiting galaxy 2MASXJ10020847+1157123
(Drake et al.~2013b\nocite{Dra13b}, 2013c\nocite{Dra13c}).

\subsubsection{Host Galaxy Metallicities}

To understand the nature of the faintest host galaxies and the hosts of SLSN, we obtained Keck LRIS
follow-up spectra of several galaxies containing CRTS supernovae after the events had significantly
faded. For the ten hosts with clear star formation present, we determined estimates for metallicity based
on the abundances of narrow lines following the method prescribed by Tremonti et al.~(2004)\nocite{Tre04}
for SDSS galaxies. This method has recently been determined to give the most accurate values, even for
extremely metal-poor galaxies \cite{Nak22}.

\begin{figure}{
\includegraphics[width=85mm]{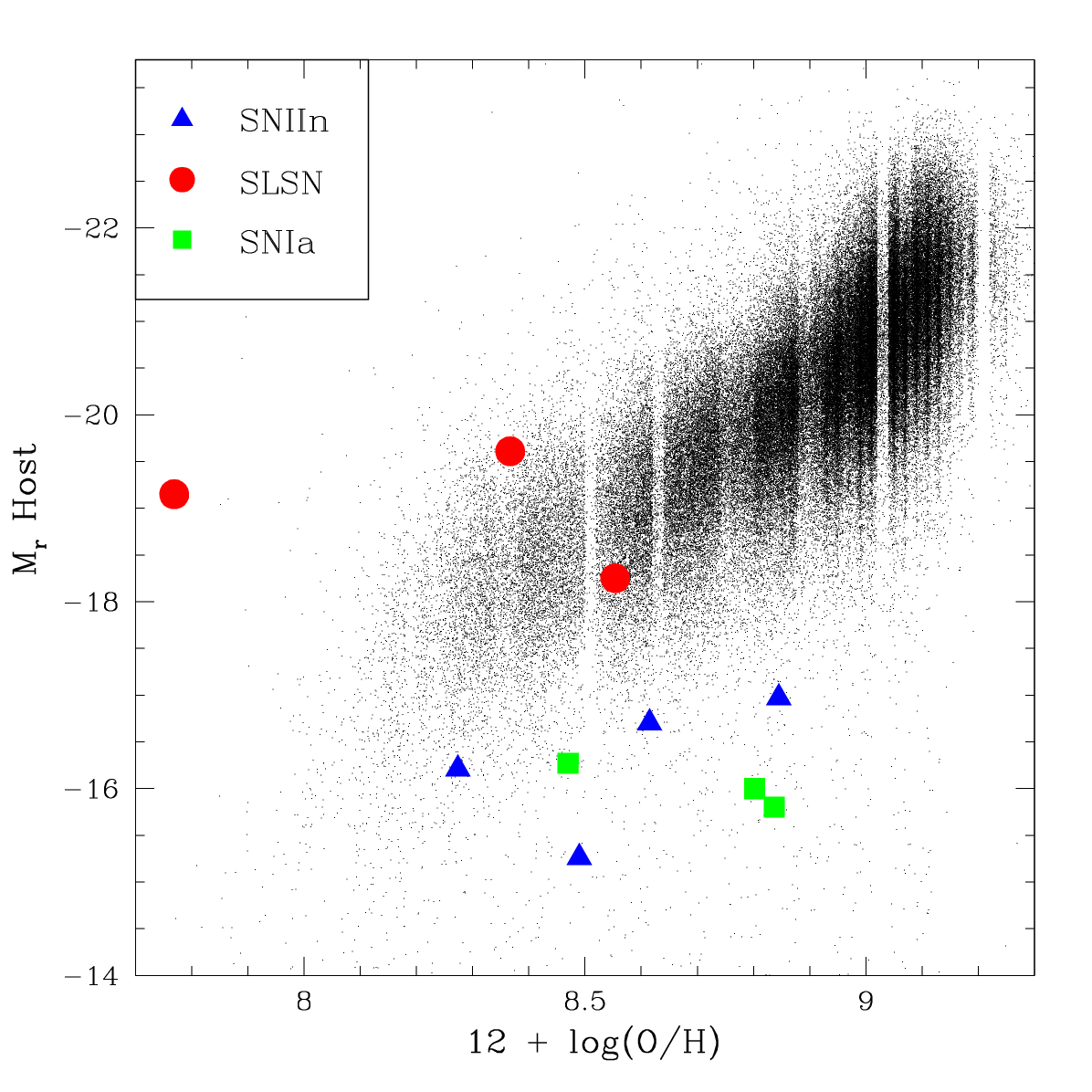}
\caption{\label{Metal}
Metallicity vs foreground extinction-corrected absolute magnitudes of CRTS supernova in faint 
host galaxies compared with values for $\sim$140,000 starforming galaxies from the SDSS DR7
MPA-JHU catalogue. The red circles mark the hosts of CRTS SLSN candidates, while the blue triangles
give type-IIn hosts, and the green boxes give type-Ia hosts.
}
}
\end{figure}

In Figure \ref{Metal}, we compare our Keck spectra-based values with the publicly available MPA-JHU 
group SDSS DR8 {\it galSpec} sample, which is based on the spectra of 140,000 star forming galaxies. 
The bands of missing bands of metallicity values within the SDSS sample are likely due to artifacts
in the data. Within our sample, the most significant outlier is the host galaxy for SN CSS081009:002151-163204.
The host has the lowest metallicity ($\rm 12+log(O/H)\sim 7.7$) and exhibits significant [OII] emission.
In general, the observed host metalicities are lower than expected if they followed the metallicity-luminosity
trend seen in the distribution of SDSS galaxies. This is different from the results of Prieto et al.~(2008)\nocite{Pri08}
which found that the supernova hosts were a good match to the SDSS sample. When compared to the less
luminosity-biased sample of supernova hosts presented by Perley et al.~(2016)\nocite{Per16}, the
values appear consistent.

\section{TDE candidates and AGN}

TDEs occur when stars are disrupted by the gravitational fields of massive black holes (MBHs) and the
material stripped in this process is accreted back onto the black hole \cite{Ree88}. 
TDEs are expected to occur in the cores of galaxies where MBHs reside. However, as AGN variability also
occurs in association with such MBHs, to find TDEs, it is necessary to separate AGN from TDEs. The
TDE detection process is further complicated by the fact that supernovae can also occur throughout
a galaxy, and supernovae are far more common than TDEs.

Past optical transient surveys initially used the presence of a nuclear event with broad He-II emission
lines at 4686\r{A} as a distinguishing feature of TDEs \cite{Gez12}. However, the spectra of TDEs are now
known to be diverse. Some events exhibit featureless blue continua, while others show broad hydrogen 
and/or helium emission lines (Hammerstein et al.~2023\nocite{Ham23}, Yao et al.~2023\nocite{Yao23}).

CRTS had limited sensitivity to the detection of TDEs since aperture photometry was used to discover
transients in CRTS, rather than image subtraction. To detect an event occurring in the core of a galaxy,
the source would often need to outshine the host galaxy (as demonstrated by the CRTS host brightness
distribution in Figure \ref{Hist}). Nevertheless, a few candidate TDEs were discovered, including
CSS100217:102913+404220 (CSS100217, Drake et al.~2011b\nocite{Dra11b}). This event occurred near a
likely AGN, making the interpretation very difficult. For example, the lightcurve and spectrum are
consistent with a SLSN-IIn, yet the spectra of type-IIn supernovae are similar to Narrow-line 
Seyfert-1 (NLSy1) AGN. The spectra of such SLSN-IIn are also now known to be similar to that of some
TDEs, as they all exhibit broad and narrow Balmer lines as well iron emission lines of similar widths.

To improve the classification of TDEs, Frederick et al.~(2021)\nocite{Fre21} recently devised a scheme 
to separate TDEs from NLSy1-associated variability. The authors discovered several TDE candidates 
and suggested that a couple of the objects that they classified as NLSy1 were similar to CSS100217. 
However, only one of the five events they selected (AT2019brs) had a lightcurve similar to that 
of CSS100217, or indeed any supernova. 
The authors then derived a binary classification scheme to separate possible TDE from AGN-related 
variability involving nine separate observables. Of these, three observables were present in all the
objects that they classified as TDEs and AGN. Selections with positive measurements of the remaining
six observables were then attributed to both cases they attributed to AGN variability and TDE. Thus,
no individual feature was able to distinguish TDEs from AGN. The authors further suggest that a couple
of their TDE candidates (AT2019avd and AT2019fdr) were unlikely to be type-IIn supernovae due to a
lack of strong P-Cygni features, thus neglecting the fact that P-Cygni features are often not seen
in the low-resolution spectra of type-IIn's (see Figure \ref{IIn}).
Additionally, more recently, Cannizzaro et al.~(2022)\nocite{Can22} suggest that CSS100217 was most
likely a TDE caused by a star in a retrograde orbit with respect to the accretion disk.

\begin{figure*}{
\includegraphics[width=85mm]{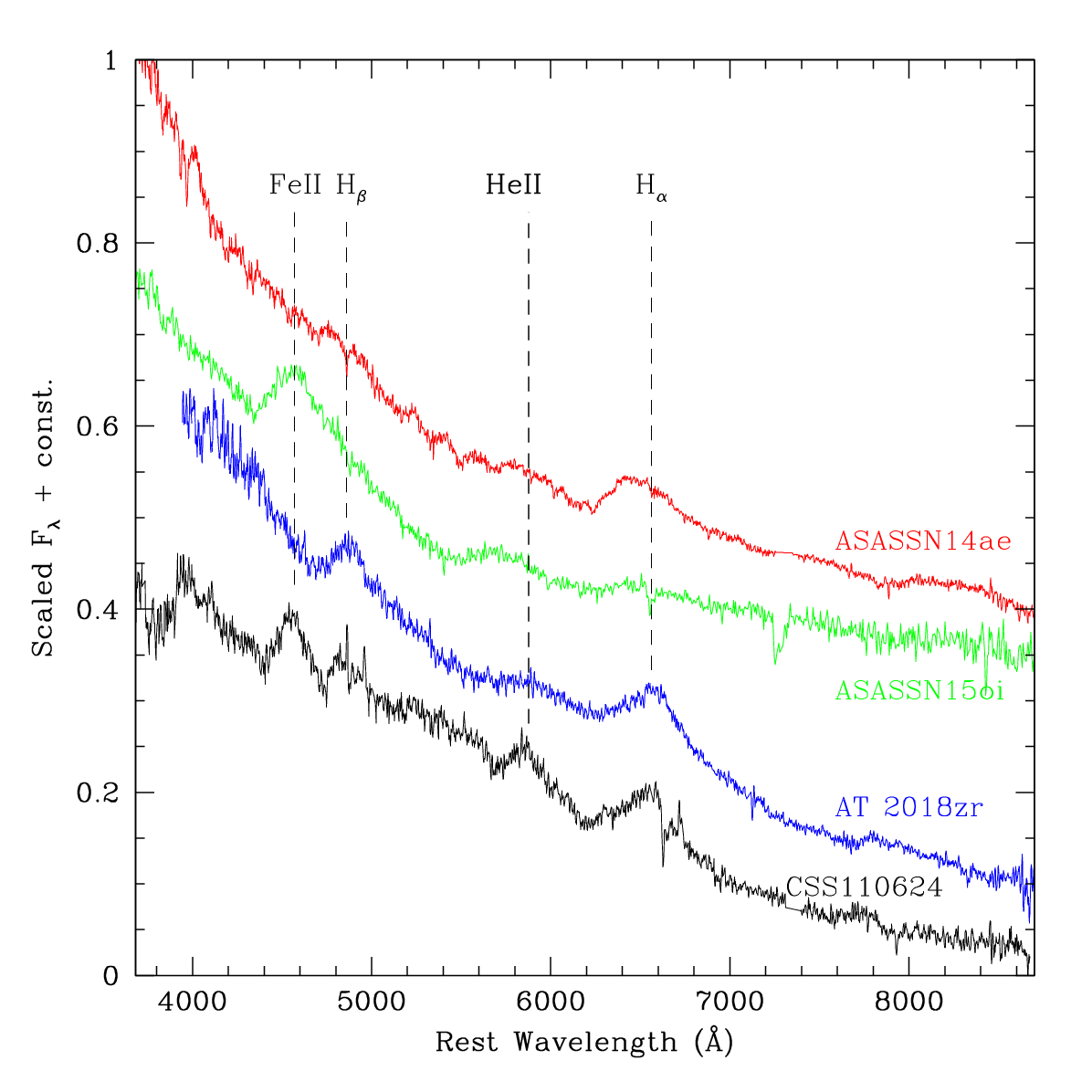}
\includegraphics[width=85mm]{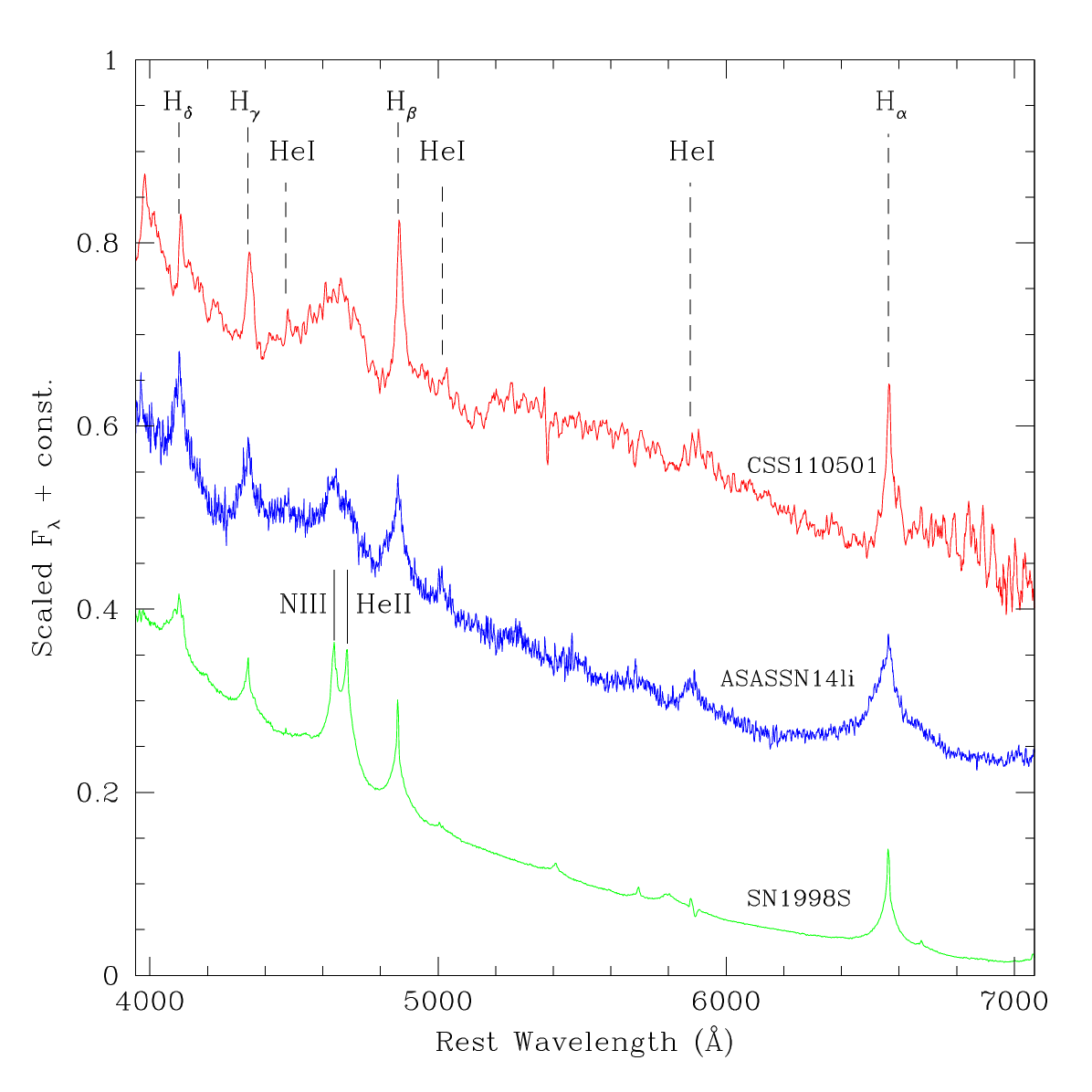}
\caption{\label{TDESpec}
Spectra of CRTS transient compared to previously known TDEs.
Left: spectra of candidate TDE CSS110624:153755+070942 (CSS110624) compared to known
TDE candidates ASASSN14ae {\protect\cite{Hol14}}, 
ASASSN15oi {\protect\cite{Hol16}},
and AT 2018zr {\protect\cite{Hol19}}.
Right: Spectra of CSS110501:144740+514105 (CSS110501, SN 2011cw Drake et al.~2011c{\protect\nocite{Dra11c}}), 
compared to TDE candidate ASASSN14li {\protect\cite{Kro16}}
and type-IIn supernova architype
SN 1998S {\protect\cite{Fra02}}.
}
}
\end{figure*}

In Figure \ref{TDESpec}, we present the spectra of CSS110624 and CSS110501 compared to previously known
TDE candidates and SN-IIn 1998S. Although CSS110624 appears to exhibit broad HeII, FeII, $\rm H_{\beta}$,
and $\rm H_{\alpha}$, as seen with some TDE candidates, the lack of colour and accurate astrometric
information makes it difficult to rule out an unusual supernova. In the case of CSS110501, the spectrum
is a close match to TDE candidate ASASSN14li's. However, the presence of likely broad Ne II and He II is
insufficient to confirm this event as more than a TDE candidate, since the SN-IIn archetype, SN 1998S,
also contained these emission features. Rather, these two candidates highlight why multi-epoch spectra
(as well as accurate photometry and astrometry) are needed to determine whether such a feature persists
at late times.

\begin{figure*}{
\includegraphics[width=85mm]{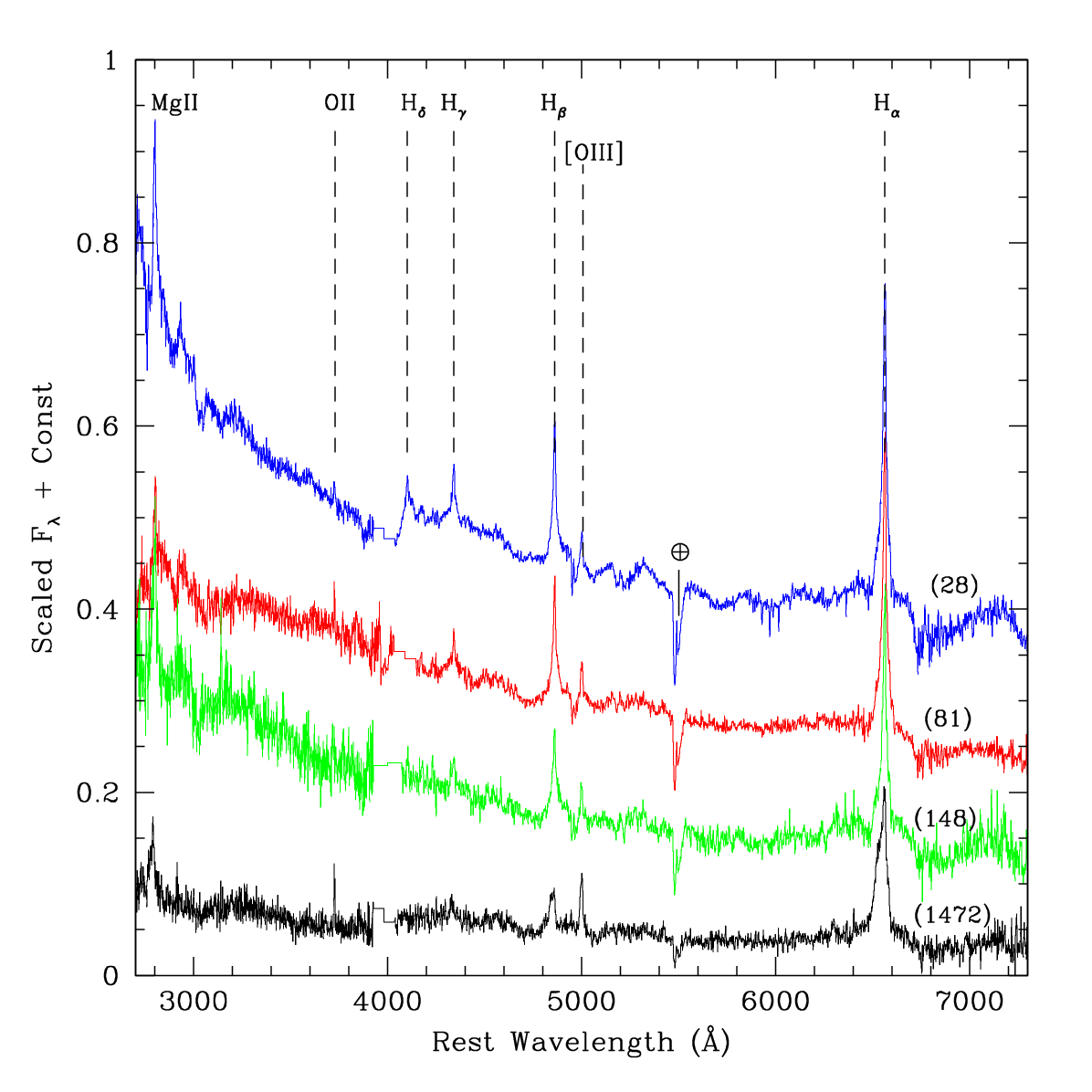}
\includegraphics[width=85mm]{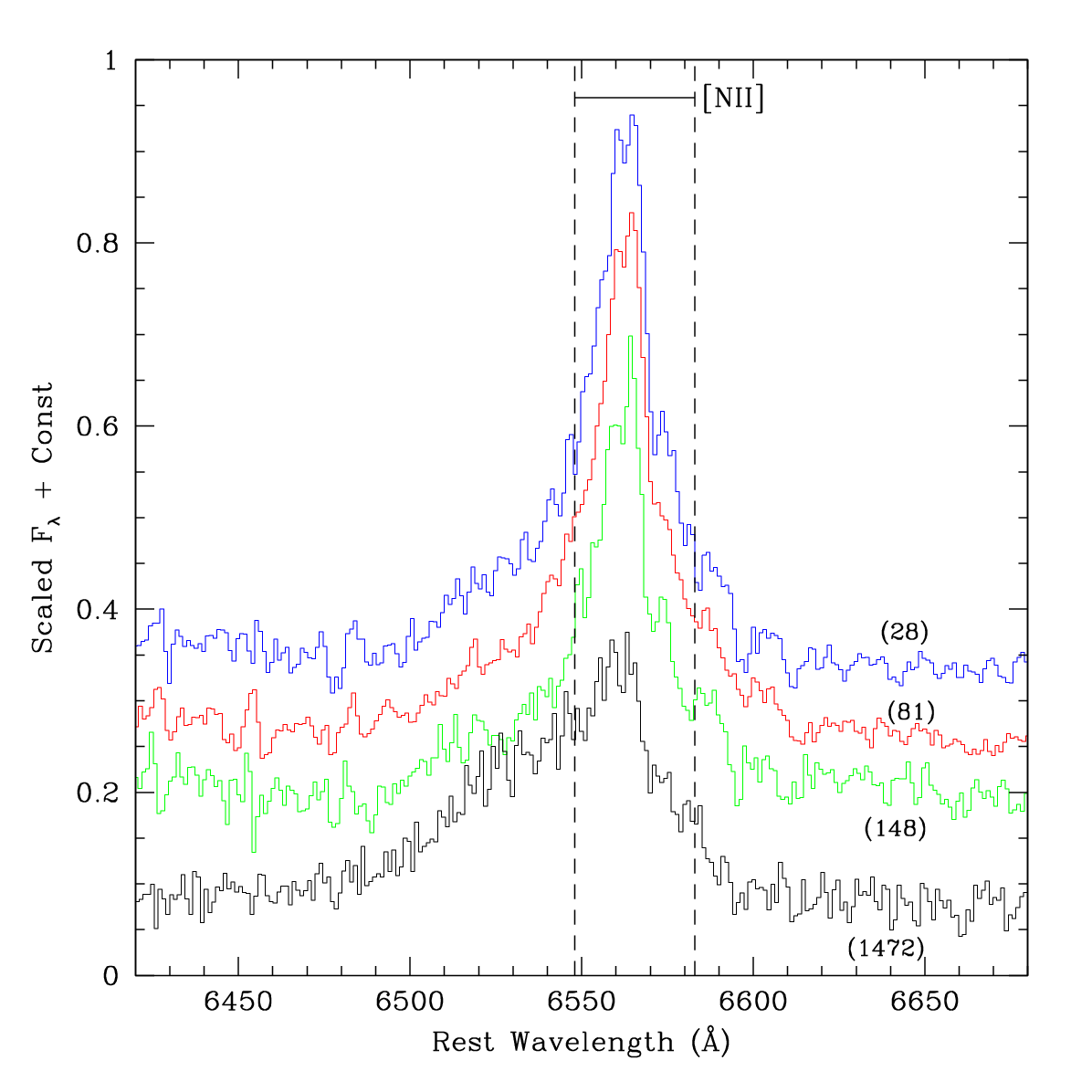}
\caption{\label{CSS150120}
Time-resolved spectra of CRTS transient CSS150120:110008+385352 (CSS150120).
In the left panel, we present spectra of the event with ages (in parentheses) 
relative to peak brightness.
In the right panel, we present the evolution of the $\rm H_\alpha$ emission.
}
}
\end{figure*}

In Figure \ref{CSS150120}, we present spectra of transient CSS150120:110008+385352 (CSS150120) 
taken over four years. The event resembles transient CSS100217 and reached a peak brightness 
of $\rm M_V = -23.6$. WISE photometry of the host taken before the transient gives $\rm W1 - W2 = 2$ 
and $\rm W1=14.85$. Based on Assef et al.~(2013)\nocite{Ass13}, this strongly suggests the presence
of an AGN. In the right panel, we can see that the shape and intensity of $\rm H_{\alpha}$ emission
changes over time with the narrow component fading while the broad, blue component remains.
Prior photometry from CRTS, as well as post-event photometry from ZTF, does not show significant 
variability. The spectra show little evidence for N II that one might expect with an NLSy1. 
We note that the moderate redshift ($z=0.386$) allows us to detect Mg II in all epochs. However,
Mg II is seen in AGN, as well as SN-IIn (Fransson et al.~2002\nocite{Fra02}, 2005\nocite{Fra05})
and the CSM of their LBV progenitors \cite{Smi19b}. 
A similar event (PS16dtm) was discovered by Petrushevska et al.~(2023)\nocite{Pet23}. 
This event appears to have occurred in an NLSy1 where it brightened by more than a magnitude in
less than 100 days then declined for more than 2000 days. Such spectroscopic variations have been
observed in changing-look QSOs. Thus, like CSS100217, CSS150120 remains difficult to classify.

\subsection{Transient Energetics}

To determine the approximate peak magnitude for each extragalactic transient, we correct
the photometry for foreground extinction based on Schlafly \& Finkbeiner~(2011)\nocite{Sch11}.
We then subtract the host fluxes to derive the peak brightness for the transient sources.
Using luminosity distances derived from the known redshifts, we determine the approximate 
peak absolute magnitude for each event. In general these estimates should be lower limits
as some additional extinction is likely present within the host galaxies.

\begin{figure}{
\includegraphics[width=85mm]{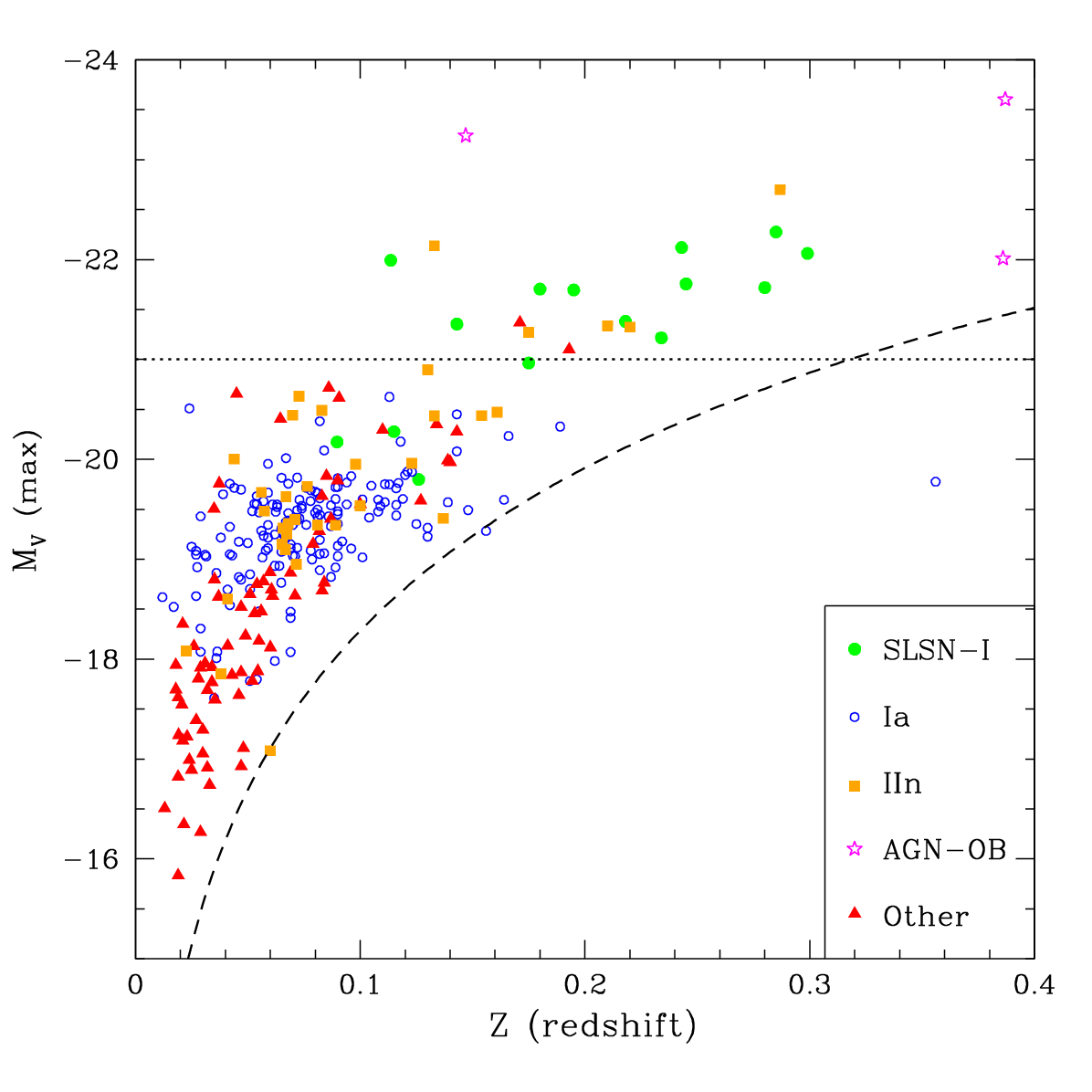}
\caption{\label{Mmax}
Peak absolute magnitudes for the spectroscopic sample.
The dashed line denotes the redshift limit expected for 
an apparent magnitude of $V = 20$. The dotted line shows
the absolute magnitude limit for events that are considered 
superluminous ($\rm M_{V} < -21$). Transient type is
denoted by the shape and colour given in the key.
}
}
\end{figure}

In Figure \ref{Mmax}, we plot the observed peak absolute magnitudes for our sample. The dashed
line shows the approximate redshift limit for a survey given a detection limit of $V=20$. This
matches the approximate limit for transient detection in CRTS.
However, in most cases we only followed the transients that were brighter than $V=19$ at peak.
The faintest supernova that we observed spectroscopically was MLS100313:085517+164511.
This event was discovered and followed near the detection limit of the MLS
observations ($\rm V_{CSS}=21.5$). 
This plot demonstrates the magnitude bias of the survey that enabled CRTS to detect superluminous
events at distances beyond $z \sim 0.32$. Similarly, faint transients, such as supernovae with
absolute magnitudes $\rm M_{V} > -16$, were only detectable at modest redshifts ($z < 0.02$).
More detailed energetics of individual events can be obtained using the CRTS data release 
photometry\footnote{http://nesssi.cacr.caltech.edu/DataRelease/}.

Lastly, Figure \ref{Mmax} also show that the most distant and luminous of the events that we obtained
spectra for are associated with host galaxies containing AGN. However, the historical lightcurves of
the hosts do not exhibit prior AGN variability, which is characterised by a damped random
walk (Kelly, Bechtold, \& Siemiginowska 2009\nocite{Kel09}, MacLeod et al.~2012\nocite{Mac12}).
These transients may be similar to that of the TDE ZTF20abrbeie \cite{Sub23}. 

\section{Discussion}

We have presented the results of spectroscopic follow-up of extragalactic transients discovered by CRTS.
Almost all the transients followed are supernovae, with the most common ($\sim 50\% $) being type-Ia's.
However, given the relatively small number of spectra that were taken (353) compared to the total number
of likely supernovae ($> 4000$) and the numbers followed by similar-sized transient surveys (e.g. PTF
with $2981$ spectra, Papadogiannakis 2019), a surprisingly large number are found to be likely SLSN-I
or SLSN-II ($\sim 31$). This is almost certainly due to a follow-up strategy that favoured luminous,
long-timescale events. 

The observational timescale bias also drives the discovery of a very large fraction (20\%) of type-IIn SN.
The most extreme example of this type of event was SN 2008iy, which was detected in CRTS photometry and
spectroscopic follow-up for more than 1800 days. In this case, we find the event likely underwent an outburst
before going supernova, similar to SN 2009ip \cite{Smi22} and other events recently identified in ZTF \cite{Str21}. 
Although SN 2008iy did not undergo a 400-day rise, as suggested by Miller et al.~(2010)\nocite{Mil10}, the
$\sim 300$ day rise time remains extremely long compared to the $50 \pm 11$ days distribution of
{\em slow-rising} type-IIn's presented by Nyholm et al.~(2020)\nocite{Nyh20}.

In addition to the common supernovae of type-IIn, many other supernovae were discovered that exhibit signs
of CSM interactions, including $\sim 24$ SN-Ia-CSM candidates. The two likely SESNe discovered by CRTS
included a so-called {\it double-humped} type-Ibn event, CSS111225. This event underwent a rebrightening
of $\sim$2.5 mags to $M_{r}= -17.5$, far in excess of the $\sim1$ magnitude observed in SN-2019oys \cite{Sol20}, 
and remained visible for more than 600 days. However, as this was only confirmed $\sim 200$ days after
the initial outburst, the exact classification remains uncertain. Similarly, SN 2009ny also appears to be
a type-Ibn but was observed at a much earlier phase.

By targeting transients that are much brighter than their host galaxies we have probed a host galaxy population
that is different from previous surveys such as LOSS that targeted bright galaxies. Indeed, many of the hosts
are also significantly fainter than galaxies spectroscopicly observed by surveys such as SDSS. However, as
only a small fraction of the host galaxies have been followed spectroscopically, future work is required to
determine how the chemical abundance and age of the hosts might affect the types of events observed.

Apart from the regular SNe, CRTS discovered a small number of transients that have spectroscopic features that
are shared with TDEs, SN-IIn, and AGN. The best examples of these are CSS100217 \cite{Dra11b} and CSS150120.
The classification of these events remains uncertain as all three types of transients can exhibit broad Balmer
and iron line complexes, X-ray emission, and have slow smoothly rising outbursts reaching $\rm M_{V} = -23$ (or brighter).
Additionally, like TDEs and AGN, where the variability occurs within the core of the galaxy, SLSN-IIn can occur
in galactic cores, possibly including even those containing AGN. For example, SLSN-II SN 2006gy occurred within $\sim 1 \arcsec$ 
of the core of NGC 1260 \cite{Ofe07} and was initially noted as being more consistent with an AGN outburst than
a supernova \cite{Pri06} 
until it was spatially resolved. However, with improved monitoring of AGN, it has become clearer that, while
outbursts do occur in AGN (Graham et al.~2017\nocite{Gra17}, Drake et al.~2018\nocite{Dra18}), they are not common.
Likewise, although many TDEs have now been observed (e.g. Hammerstein et al.~2021\nocite{Ham21}), they are still
rare compared to supernovae. Future work following the analysis of Frederick et al.~(2021)\nocite{Fre21} may
enable a clearer separation of the small number of events with overlapping features.

The increased photometric monitoring of millions of galaxies by modern transient surveys has resulted in a significant
increase in the rate of discovery of extragalactic transients. This increase limits spectroscopic confirmation to a
small fraction of the events being discovered. However, event classification has made large advances through machine
learning (e.g. Mahabal et al.~2019\nocite{Mah19}, Smith et al.~2019\nocite{Smi19b}, Forster et al.~2021\nocite{For21})
and is further aided using fast low-resolution spectra \cite{Bla18}. 
Yet extracting the most science, when most transients are too faint to spectroscopically confirm, remains a particularly
important issue for deeper surveys such as Rubin's LSST \cite{Ive19}. To some extent this may be alleviated by future
wide-field telescopes such as the WST \cite{Mai24}. 

\section*{Acknowledgments}

CRTS, CSDR1 and CSDR2 are supported by the U.S.~National Science Foundation under grant NSF grants AST-1313422,
AST-1413600, AST-1518308, and AST-1749235.  The CSS survey is funded by the National Aeronautics and Space Administration
under Grant No. NNG05GF22G issued through the Science Mission Directorate Near-Earth Objects Observations Program.
SGD acknowledges generous support from the Ajax Foundation.
AJD and MC acknowledge partial support by CONICYT's PCI program through grant DPI20140066. MC is additionally supported
by ANID's FONDECYT Regular grants \#1171273 and 1231637; ANID's Millennium Science Initiative through grants
ICN12\textunderscore 009 and AIM23-0001, awarded to the Millennium Institute of Astrophysics (MAS); and ANID's
Basal project FB210003. JLP acknowledges support from ANID, Millenium Science Initiative, AIM23-0001.
This work has made use of spectra obtained via WISeREP - https://www.wiserep.org.

The Pan-STARRS1 Surveys (PS1) and the PS1 public science archive have been made possible through contributions 
by the Institute for Astronomy, the University of Hawaii, the Pan-STARRS Project Office, the Max-Planck Society
and its participating institutes, the Max Planck Institute for Astronomy, Heidelberg and the Max Planck Institute 
for Extraterrestrial Physics, Garching, The Johns Hopkins University, Durham University, the University of Edinburgh, 
the Queen's University Belfast, the Harvard-Smithsonian Center for Astrophysics, the Las Cumbres Observatory Global 
Telescope Network Incorporated, the National Central University of Taiwan, the Space Telescope Science Institute, 
the National Aeronautics and Space Administration under Grant No. NNX08AR22G issued through the Planetary Science 
Division of the NASA Science Mission Directorate, the National Science Foundation Grant No. AST–1238877, the 
University of Maryland, Eotvos Lorand University (ELTE), the Los Alamos National Laboratory, and the Gordon and 
Betty Moore Foundation. The Sloan Digital Sky Survey (SDSS) is a joint project of The University of
Chicago, Fermilab, the Institute for Advanced Study, the Japan Participation
Group, The Johns Hopkins University, the Los Alamos National Laboratory, the
Max-Planck-Institute for Astronomy (MPIA), the Max-Planck-Institute for
Astrophysics (MPA), New Mexico State University, University of Pittsburgh,
Princeton University, the United States Naval Observatory, and the
University of Washington. Funding for the project has been provided by the
Alfred P. Sloan Foundation, the Participating Institutions, the National
Aeronautics and Space Administration, the National Science Foundation, the
U.S. Department of Energy, the Japanese Monbukagakusho, and the Max Planck
Society. www.sdss.org is a winner of the Griffith Observatory's Star Award

\section*{Data Availablity}

The spectroscopic data used in this paper will be made publicly available from the CRTS release 
website (http://nesssi.cacr.caltech.edu/DataRelease/) upon publication and is available 
request to the corresponding author. Photometric measurements for all CRTS transients is currently public
through the same address. Comparison spectra and PanSTARRS photometry are publicly available via 
WISeREP (https://www.wiserep.org) and MAST (https://archive.stsci.edu/panstarrs/), respectively. 
SDSS galaxy line measurements are public available via the MPA/JHU catalog (https://wwwmpa.mpa-garching.mpg.de/SDSS/).

\clearpage

\end{document}